\UseRawInputEncoding
\documentclass[showpacs,preprintnumbers,amsmath,amssymb,aps, prd]{revtex4}
\usepackage{graphicx}
\usepackage{grffile}
\usepackage{caption}
\usepackage{subcaption}
\usepackage{float}
\usepackage{xcolor}
\usepackage{dcolumn}
\usepackage{bm}
\begin{document}
\title{Superradiance scattering off rotating Simpson-Visser black hole
and its shadow in the non-commutative setting}
\author{Sohan Kumar Jha}
\affiliation{Chandernagore College, Chandernagore, Hooghly, West
Bengal, India}
\author{Anisur Rahaman}
\email{anisur.associates@iucaa.in; manisurn@gmail.com
(Corresponding Author)} \affiliation{Durgapur Govt. College,
Durgapur, Burdwan - 713214, West Bengal, India}

\date{\today}
\begin{abstract}
\begin{center}
Abstract
\end{center}
We consider non-commutating Simpson-Visser spacetime and study
the superradiance phenomena and the shadow cast by the back hole
associated with this modified spacetime. We extensively study the
different aspects of the black hole associated with the metric
endowed with the corrections linked with non-commutative
properties of spacetime. We study the superradiance
effect, deviation of shape, size of the ergosphere, and
the shadow of black
hole in this extended situation and look into their variation
taking different values Simpson-Visser parameter $\ell$ and
non-commutative parameter $b$. We have made an attempt to
constrain the parameter $\ell$ using the data available from the EHT collaboration for $M87^*$ black hole.
Our study reveals that black holes are associated with non-commutative
Simpson-Visser spacetime may be a suitable candidate for an
astrophysical black hole.
\end{abstract}
\maketitle
\section{Introduction}
Black holes are the most intriguing prediction of the general
theory of relativity (GTR). The existence of black holes in nature
has now been evident from the theoretical formulation with 
adequate evidence from the astronomical observation. The recent
success of capturing the shadow of the $M87^*$ supermassive black
hole, followed by the success of capturing the shadow of the
$SgrA^*$ black hole in our galaxy, and of course the triumphs
of gravitational wave detection, provide the most compelling proof
of their existence. When light approaches the vicinity of the
black hole various thrilling optical phenomena have been found to
set up. The lensing effect (strong and weak), formation of the shadow, and superradiance scattering are remarkable in this environment.
The study of the optical effect in the vicinity of a black hole
was started in the long past and it was developed through
the efforts of different scientists in due course. Superradiance
is a fascinating optical phenomenon connected to the scattering of
electromagnetic waves. It has been established that in a
gravitational system, the scattering of radiation off absorbing
rotating object produces electromagnetic waves with amplitude
larger than that of the incident one when a specific condition is
maintained between the frequency of the incident radiation and
angular velocity of the rotating gravitational system \cite{ZEL0,
ZEL1}. In 1971, Zel'dovich in his seminal work showed that the scattering of radiation off rotating absorbing surfaces may result
in electromagnetic waves with a larger amplitude if $ \omega <
m\Omega$ is maintained. Here $\omega$ is the frequency of the
incident monochromatic radiation, $m$ is the azimuthal number with
respect to the rotation axis and $\Omega$ is the angular velocity
of the rotating gravitational system \cite{ZEL0, ZEL1}. The
contemporary contributions related to superradiance \cite{STRO1,
STRO2, TEUK, PRESS} made these astounding astronomical phenomena a
tempting field of research. The lecture notes \cite{REVIEW} and
the references therein contain an excellent review on
superradiance phenomena. This phenomenon is associated with
numerous classical problems that include stimulated or spontaneous
energy emigration, e.g. the Vavilov-Cherenkov effect, the
anomalous Doppler effect, etc. The discovery of the black hole
evaporation was well understood from the studies of black hole
superradiance \cite{HAW}. Interest in the study of black hole
superradiance has recently been revived in different areas,
including astrophysics and high-energy physics via the
gauge/gravity duality along with fundamental issues in the general
theory of relativity (GTR). Superradiant instabilities can be used
to constrain the mass of ultra-light degrees of freedom
\cite{INST00, INST0, INST1, INST2}, with important applications to
dark-matter searches. The black hole superradiance is also
associated with the existence of new asymptotically flat hairy
black-hole solutions \cite{HAIR} and with phase transitions
between spinning and/or charged black objects and asymptotically
anti-de Sitter (AdS) spacetime \cite{ADS0, ADS1, ADS2} or in
higher dimensions \cite{MSHO}. Furthermore, the knowledge of
superradiance is instrumental in describing the stability of black
holes and determining the fate of the gravitational collapse in
confining geometries \cite{ADS0}

Another optical phenomenon of black holes which is of great
interest is the black hole shadow. A two-dimensional darkish
location in the celestial sphere prompted by the strong gravity of
the black hole is termed as shadow. It was initially pointed out
and examined thoroughly by Synge in 1966 for a Schwarzschild black
hole \cite{SYNGE}. Luminet \cite{JP} determined the radius of
the shadow in the following years. The shadow of a non-rotating black hole has circular geometrical
structure, while the shadow of a rotating black hole elongates in
the direction of the rotating axis due to the dragging effect of
spacetime \cite{BARDEEN, CHANDRASEKHAR}. Hioki and Maeda \cite{KH}
proposed two observables based on the feature that the boundary of
the Kerr shadow would match with the astronomical observations.
One of the observables roughly describes the deformation of its shape from a
reference circle and the other describes the size of the shadow.
The deviation from circularity $\Delta C$ can also be determined
using the method given in \cite{TJDP}. These observables are
extremely serviceable in testing astronomical phenomena and
constraining the free parameters linked with the theories of
gravity. When the quantum effect was included it was argued that the rotational superradiance would count as a spontaneous process
where rotating bodies including black holes would slow down by the
emission of radiation.

The black hole itself is one of the fascinating predictions of the
GTR. After the tremendous success of EHT collaboration \cite{KA1,
KA2, KA3, KA4, KA5, KA6} in capturing the shadow of the
supermassive black hole $M87^*$ at the center of the nearby galaxy
Messier $87$ and the shadow of $SgrA^*$ in our Milky Way galaxy, the study of optical phenomena in the vicinity of black holes have
attracted increased interest \cite{DENTON, MENG, KDADI, KDADI1,
KDADI2, KINO}. To clarify the uncertainties that occasionally
appear in various physical observables, numerous modified
theories of gravity were developed in addition to the conventional
theory of gravity over time. These modified theories have also
rendered their service toward precise explanations for
observations which became feasible with the help of instruments of
modern sophistication.

The inclusion of a quantum correction is clearly a viable
direction for improving the standard theory of gravity. It is of
importance and considerable significance since a fully developed
form of quantum gravity is still not at hand. In recent times,
therefore, the correction associated with taking in the impact of
the Planck scale is immensely cherished. Modifying the theory of
gravity with the effects of non-commutative spacetime is one
technique to account for the Planck effect, which is seen as
having major significance in contemporary literature. However,
here lay some fundamental problems concerning the principle with
which physical theories are built. It is none other than the
maintenance of Lorentz invariance. Over the years physicists are
engaged in examining the effects of the non-commutative character
of spacetime in various fields. In recent times, significant
development has been made using the effect of amending the
non-commutative aspect of spacetime supposed to occur in the
vicinity of the Planck scale \cite{ZABONON, NONCOM, JINNON,
ANINON, GITNON}. Black hole physics is a fertile area for
investigating the consequences of modified/extended spacetime
complemented by the spacetime non-commutativity which is believed
to happen in the Planck scale. There are a number of techniques to
incorporate the non-commutative aspect of spacetime in the
theories of gravity \cite{ PNICO, PAS, PAS1, SMEL, HARI}.

The principle of Lorentz invariance is essential to the
formulation of the general theory of relativity (GTR) and the
Standard Model (SM) of particle Physics. The formulation of the
standard model and the GTR relies entirely on the principle of
Lorentz invariance. The GTR does not take into account the quantum
properties of particles and SM, on the other hand, neglects all
gravitational effects of particles. Within the vicinity of the
Planck scale, one cannot neglect gravitational interactions
between particles, and hence it is essential to merge SM with GTR
in a single theory to have fruitful results. It is indeed
available from the quantum gravity concept which is viable at the
Planck scale. Unfortunately, the theory of quantum gravity is not
available in a mature stage. At this scale, spacetime may exhibit its non-commutative nature.
Therefore, the implementation of the non-commutative character of
spacetime can be viewed indirectly as a supplement to quantum
gravitational effects. However, \cite{DM} due to the presence of the
real and anti-symmetric $\theta_{\mu\nu}$ within the basic
formulation of non-commutative extension of spacetime $[x_\mu,
x_\nu] = i\theta_{\mu\nu}$, violation of Lorentz symmetry is
inherent within the non-commutative theories. In the
papers \cite{SMA, SMA1, SMA2} Smailagic et al. and in \cite{NICO1}
Nicolini et al. simulated non-commutative spacetime in an
intriguing manner by invoking the ingenious coordinate coherent
state formalism that kept the Lorentz symmetry preserved. In this
extension, we tried to provide a faithful and decent framework
that takes into account the non-commutative properties of spacetime.
Therefore, Lorentzian symmetry is preserved in this extended
framework because the way in which the non-commutative aspects of
space-time is amended here is not in direct conflict with
Lorentz symmetry

The main idea behind the modification of spacetime is to replace
the point mass with a distribution of mass. Through the
replacement of the point-like matter source (likely to describe by
delta function) by a distribution of mass by Gaussian distribution
or Lagrangian distribution, amendment of NC spacetime came into the
literature. In the paper \cite{NICO1}, Gaussian distribution, and
in the paper \cite{NOZARI}, Lorentzian mass distribution was used
to implement the non-commutative effect. An interesting extension
concerning the thermodynamic similarity between
Reissner-Nordstr$\ddot{o}$m black hole and the NC Schwarzschild
black hole has been made in the paper \cite{KIM}. The
thermodynamical aspects of NC black holes have been investigated
by considering the tunneling formalism in the papers \cite{KNM,
RABIN, MEHDI, MEHDII, SUFI,ISLAM, GUPTA}. The impact of
non-commutative spacetime in cosmology has been studied
extensively in the papers \cite{PINTO, MARCOL, ELENA, MOHSEN,
ABD}. In \cite{LIANG}, with the aid of taking the mass to be a
Lorentzian smeared mass distribution the thermodynamic properties
of NC BTZ black holes have been studied. So far we have found that
non-commutative spacetime can be implemented by replacing  the
point-like source of matter designated by the Dirac delta function with a smeared distribution of matter. In papers \cite{NOZARI, NICO1} Gaussian and Lorentzian distributions
are used to comprise the idea of non-commutative spacetime.
Keeping it in view, in this manuscript, we have introduced the
non-commutative setting into the Simpson-Visser spacetime with the aid of Lorentzian distribution as it was used in \cite{NOZARI}.
There are several other ways, of course, in the center where
a quantum correction was made but Lorentz symmetry was not possible
to maintain. An instance in this regard is the implementation of
the bumblebee field to encompass quantum effect \cite{EMS1, EMS2,
EMS3, EMS4, EMS5, EMS6, EMS7, EMS8, EMS9, EMS10, EMS11,
EMS12,EMS13, EMS14, EMS15, EMS16, EMS17, RB, GVL1, GVL2, GVL3,
GVL4, STRING1, STRING2, STRING3, STRING4, DC, DC1, ESM1, ESM2,
DING, RC, ARS}

In this bid, we will concentrate on probing the impact of the
quantum gravity effect due to the non-commutative aspect of
spacetime on the superradiance scattering off black holes and the
shadow cast by the black holes. Thus, non-commutative correction
of spacetime has been made in the Simpson-Visser spacetime
background and an attempt has been made to carry out an
investigation in order to observe its effect on superradiance
phenomena and the shadow corresponding to the black hole in this
modified spacetime background. This new frame will allow us to study
systematically the amount of correction on superradiance and the
black hole shadow. The Simpson-Visser black hole falls in the class of the regular black hole model.
The existence of a singularity, by its definition, suggests that
spacetime ceases to exist, marking a breakdown of the laws of
physics. Any reasonably extreme condition, which may exist at the
singularity implies that one ought to trust quantum gravity that
is anticipated to resolve this \cite{WHEELER}. In the absence of
any definite quantum gravity, necessary attention shifted to
regular models which might permit one to grasp the interior of the
black hole and resolve the singularity issue. The idea of regular
models was pioneered by Sakharov \cite{SAKHAROV} and Gliner
\cite{GLINER}, which implies that one may eliminate singularities by considering the presence of matter. Motivated by
Sakharov \cite{SAKHAROV} and Gliner \cite{GLINER}, Bardeen
proposed a regular black hole that showed the promising existence
of horizons \cite{BARDEEN}. The Bardeen black hole close to the
origin acts just like the de Sitter spacetime, whereas for large
$r$, it resembles the Schwarzschild black hole. The non-singular
black hole proposed by Hayward is also intriguing. Lately, another
attention-grabbing spherically symmetric regular black hole was
projected by Simpson and Visser \cite{SIMP1, SIMP2, SIMP3}. There
is a free parameter having the dimension of length introduced
there to cause repulsion in order to avoid singularity and it is
accountable for the regularization of the metric at r =0.
Therefore the non-commutative extension of the Simpson-Visser back
hole and study of superradiance phenomena and the shadow with this
extended setting is of interest. What light the gathered knowledge
from the $\mathrm{M}87^{*}$ data can shed on this modified
framework that would also be an attention-grabbing extension.
Through our analysis, we will make an attempt to constrain the parameter $(\ell)$ from the data of EHT collaboration
concerning the shadow of the $\mathrm{M}87^{*}$ black hole. The
important information about the shadow which is available now is
as follows. It has an angular diameter of $42\pm 3\mu$ with the
deviation from circularity $\Delta C= 0.1$ and axial ratio
$\approx \frac{4}{3}$. These are the experimental inputs we will
use to constrain the free parameter involved in this modified
theory of gravity.

The manuscript is organized as follows. In Sec. II, we briefly
describe the rotating Simpson-Visser metric and the amendment
of the non-commutative aspect of spacetime in it. In Sect. III, we
study the geometrical aspects concerning the horizon and
ergosphere of this modified metric. Sec. IV and its two
subsections are devoted to describe superradiance scattering off
black hole corresponding to this augmented metric in detail. In
Sec. V we describe the photon orbit and shadow cast by the
black hole associated with this modified metric. In Sec. VI, an
attempt has been made to constrain the parameters from the
observation of the EHT collaboration conserving the
$\mathrm{M}87^{*}$ black hole. Sec. VII contains a summary
and conclusion of the work.

\section{Simpson-Visser black holes}\label{Sec2}
Simpson and Visser proposed a very simple spherically symmetric
and static spacetime family in \cite{SIMP1}. It is defined by the
metric
\begin{equation}
ds^2=-(1-\frac{2M}{\sqrt{r^2+\ell^2}})dt^2
+(1-\frac{2M}{\sqrt{r^2+\ell^2}})^{-1} + (r^2+l^2)(d\theta^2+
sin^2\theta d\phi^2)
\end{equation}
here $M \ge 0$ is the ADM mass and $\ell$ is a parameter having a
dimension of length. It is theoretically appealing since it
provides a unique description of regular black holes and wormholes
through smooth interpolation of the possibilities generated in
terms of parameter $\ell$ that drives the regularization of the
central singularity. It describes a two-way, traversable wormhole
for $\ell > 2M$ and a one-way wormhole with a null throat for
$\ell = 2M$. It is a regular black hole where singularity is
replaced by a bounce to a different universe for $\ell < 2M$. The
bounce takes place through a spacelike throat shielded by an event
horizon and it is christened as black-bounce in \cite{SIMP1} and
hidden wormhole in \cite{SIMP4}. Its rotational form was recently
shown as well in \cite{MAZZA}. We consider that rotating
Simpson-Visser black hole developed in \cite{MAZZA} which
contains additional parameter $\ell$ apart from mass $M$ and
angular momentum $a$, which in Boyer-Lindquist coordinates, is
given by \cite{MAZZA}
\begin{eqnarray}\label{metric1}
ds^2 &=& -\left(1-\frac{2M \sqrt{r^2+\ell^2}}{\Sigma}\right)dt^2
+ \frac{\Sigma}{\Delta_{sv}} dr^2+ \Sigma d\theta^2
-\frac{4Ma \sin^2\theta\sqrt{r^2+\ell^2}}{\Sigma} dtd\phi \nonumber
\\ && + \frac{A\sin^2\theta~}{\Sigma} d\phi^2,
\end{eqnarray}
where $\Sigma = r^2+\ell^2 +a^2\cos^2\theta$, $\Delta_{sv} =
r^2+\ell^2+a^2-2M\sqrt{r^2+\ell^2}$, and $A =(r^2+\ell^2
+a^2)^2-\Delta_{sv} a^2\sin^2\theta$. The additional parameter
$\ell$ is what deviates the metric~(\ref{metric1})
from Kerr black hole and one gets back the Kerr black hole
a special case ($\ell=0$). When $a =0$, we get the spherical
Simpson--Visser metric\cite{SIMP1, SIMP2, SIMP3}.
When both the parameters $a$ and $\ell$ are zero, it reduces to the
familiar Schwarzschild solution. The non-negative parameter $\ell$
with dimensions of length causes a repulsive force that avoids
singularity at $r=0$. The metric~(\ref{metric1}), under the
transformation $r\to-r$, is symmetric. It is evident that the
surface $r=0$ is a regular surface of finite size for
$\ell\ne 0$.
Under the condition $\ell\ne0$ the observer can easily cross
the surface $r=0$~\cite{MAZZA}. Depending on the value of the
parameter $\ell$ metric~(\ref{metric1}) is either a regular
black hole or a traversable worm-hole~\cite{MAZZA}

\subsection{Ratting Simpson-Visser black hole with the non-nominative setting}
Let us now describe in short the consolidation of the impact of
non-commutativity. The non-commutative spacetime within the theory
of gravity has been the subject of several interests \cite{NICO,
ZABO}. Albeit an ideal non-commutative augmentation of the
standard theory of gravity has not yet been accessible, it
necessarily requires getting the impact of non-commutativity in
the edge of the commutative theory of standard general relativity
because the studies of different aspects of black holes with the
non-commutative effects have attracted huge attention. In recent
times, in the papers \cite{NICO, ZABO, SMA, SMA1, NICO1} attempts
have been made to augment Schwarzschild's black hole solution with
the non-commutative effect. In this respect
generalization of quantum field theory by non-commutativity based
on coordinate coherent state formalism has been found to be
instrumental \cite{SMA, SMA1, NICO1, NOZARI}. In this formalism,
it was considered that the proper mass $M$ was not localized at a
point. It was distributed throughout a region of linear size. The
imputation of this argument needs the replacement of the position
Dirac-delta function which describes a point-like structure, with
the suitable function describing smeared structures. The Gaussian
distribution \cite{NICO1} function, as well as Lorentzian
distribution \cite{NOZARI} fits well in this context. Thus, by
modifying the point-like mass density described in terms of the
Dirac delta function by a Gaussian distribution or alternatively
by a Lorentzian distribution non-commutativity can be amended in
the theory in a decent and acceptable manner. Note that in reality, the description of matter should not be a Dirac-delta
distribution. It would be better described by a Gaussian or
Lorentzian distribution or some other type of distribution that
turn into a Dirac-delta function when the width of the
distribution function approaches a vanishing limit.

In order to include the non-commutative effect we chose a Lorentzian
distribution for the mass density of the black hole following the
development in the papers \cite{NOZARI, ANACLETO}
\begin{equation}
\rho_{b}=\frac{\sqrt{b} M}{\pi^{3 / 2}\left(\pi b+r^{2}\right)^{2}}.
\end{equation}
Here $M$ is the total mass distributed throughout a region with a
linear size $\sqrt{b}$ where $b$ is the strength of non-commutative
character of spacetime. With the Lorentzian type smeared matter
distribution function \cite{ANACLETO}, it can be shown that
\begin{equation}
\mathcal{M}_{b}=\int_{0}^{r} \rho_{b}(r) 4 \pi r^{2} d r =\frac{2
M}{\pi}\left(\tan ^{-1}\left(\frac{r}{\sqrt{\pi b}}\right)
-\frac{\sqrt{\pi b} r}{\pi b+r^{2}}\right) \approx -\frac{4
\sqrt{b} M}{\sqrt{\pi} r}+M+\mathcal{O}\left(b^{3 / 2}\right).
\end{equation}
It clearly shows that the Mass turns into a point-like shape when
the spread of the distribution approaches towards zero: $\lim_{b
\to 0 }M_b = M $.

When the metric for the Simpson-Visser black hole is augmented
with the noncommutative effect, it reads
\begin{eqnarray}\label{FINAL}
ds^2 &=& -\left(1-\frac{2M_{b} \sqrt{r^2+\ell^2}}{\Sigma}\right) dt^2
+ \frac{\Sigma}{\Delta} dr^2+ \Sigma d\theta^2
-\frac{4M_{b}a \sin^2\theta\sqrt{r^2+\ell^2}}{\Sigma} dtd\phi \nonumber \\
&& + \frac{A\sin^2\theta~}{\Sigma} d\phi^2,
\end{eqnarray}
where $\Sigma = r^2+\ell^2 +a^2\cos^2\theta$, $\Delta =
r^2+\ell^2+a^2-2M_{b}\sqrt{r^2+\ell^2}$, and $A =(r^2+\ell^2
+a^2)^2-\Delta a^2\sin^2\theta$. It transpires that if it is
augmented with the non-commutative effect, the surface $r=0$ no
longer remains a regular surface. The parameter $b$ is connected
with the non-commutative aspect of spacetime. If we set $b=0$, we
get the metric for the rotating (commutating) Simpson-Visser black
hole. On the other hand, if $\ell$ is set to a vanishing value,
we get a non-commutative Kerr black hole and if $\ell \rightarrow
0$ along with $b \rightarrow 0$, the metric (\ref{FINAL}) turns
into the usual Kerr metric.
\section{Geometry concerning Horizon and ergo-sphere}
We are now in a position to begin our investigation with the
metric developed in equation (\ref{FINAL}). We get the expressions for
Event horizon and Cauchy horizon by setting
$\Delta=0$. Let us
now plot $\Delta$ for different values of $b$ and $\ell$ to
observe the nature of singularity distinctly in Figs. (1, 2).

\begin{figure}[H]
\centering
\begin{subfigure}{.5\textwidth}
\centering
\includegraphics[width=.7\linewidth]{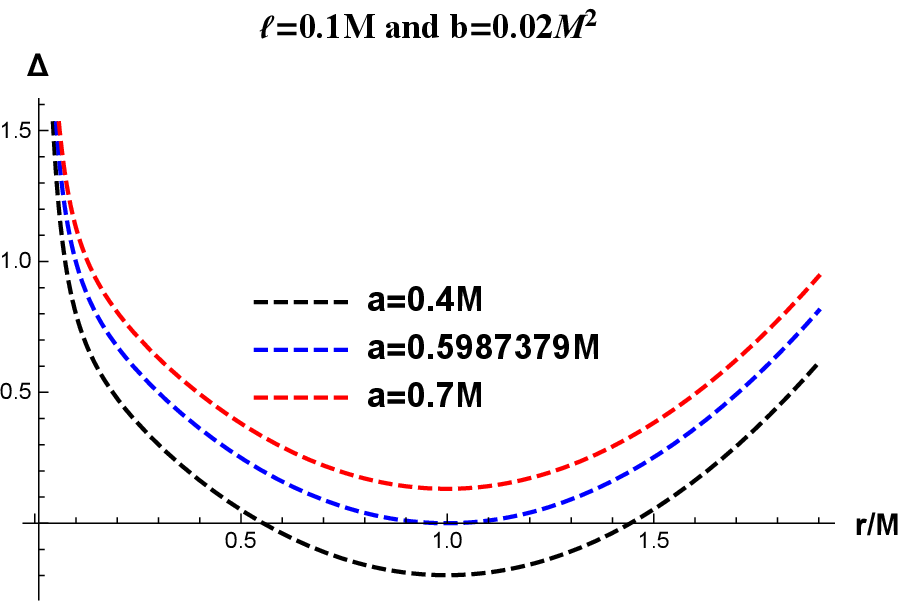}
\end{subfigure}%
\begin{subfigure}{.5\textwidth}
\centering
\includegraphics[width=.7\linewidth]{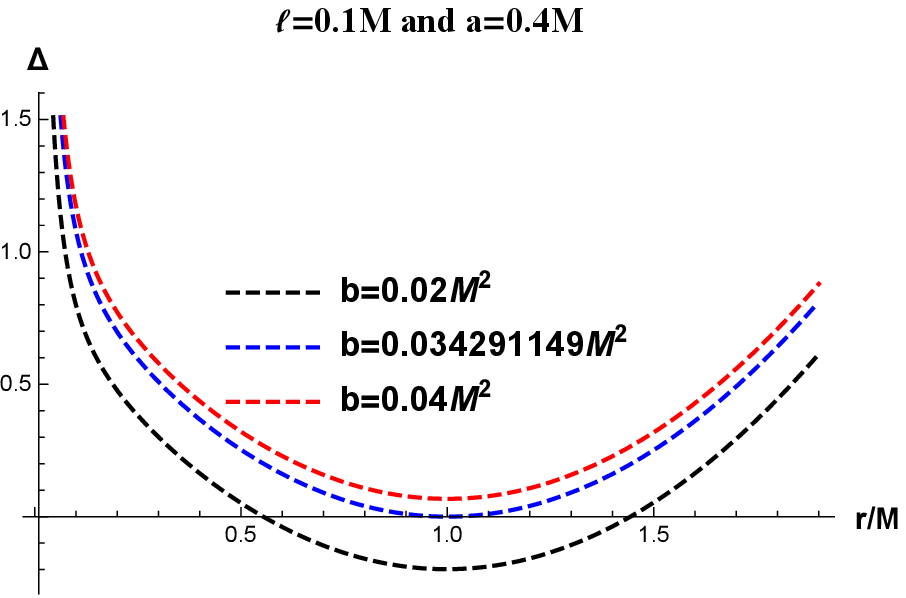}
\end{subfigure}
\caption{The left one gives the variation of $\Delta$ for various
values of $a$ with $b=0.02M^{2}$ and $\ell=0.1M$, and the right one
gives variation for various values of $b$ with $a=0.4M$ and
$\ell=0.1M$.} \label{fig:test}
\end{figure}

\begin{figure}[H]
\centering
\begin{subfigure}{.5\textwidth}
\centering
\includegraphics[width=.7\linewidth]{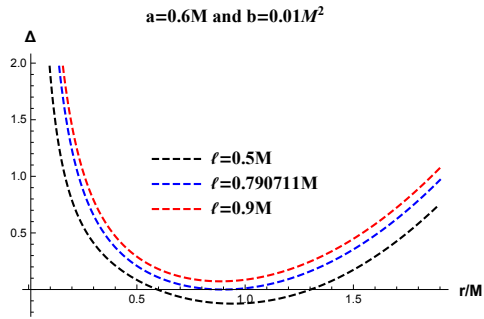}
\end{subfigure}
\caption{It gives variation of $\Delta$for various values
of $\ell$ with $b=0.01M^{2}$ and $a=0.6M$. }
\label{fig:test}
\end{figure}
The plots in Figs. (1, 2) show some critical values of the
parameter involved in the metric. We observe that there exists
critical values of $a$, for fixed values of $b$ and $\ell$.
Similarly, we get critical values of $b$ for fixed values of $a$
and $\ell$ as well as critical values of $\ell$ for fixed
values of $b$ and $a$. The critical values of $a$, $b$, and $\ell$
are designated by $a_c, b_c$, and $\ell_c$ respectively. In these
cases, $\Delta=0$ has only one root. For $a < a_{c}$ we have black
hole, however, for $a
> a_{c}$ we have naked singularity. Similarly for $b < b_{c}$ we
have black hole, but for $b > b_{c}$ we have naked singularity,
and for $\ell < \ell_{c}$ signifies the existence of black hole, whereas $\ell
> \ell_{c}$ represents the naked singularity. Numerical
computation shows that we have $a_{c}=0.5987379M$ for
$b=0.02M^{2}$ and $\ell=0.1M$. Similarly, for $a=0.4M$ and
$\ell=0.1M$ we have $b_{c}=0.034291149M^{2}$ and for $a=0.6M$ and
$b=0.01M^{2}$ we find $\ell_{c}=0.790711M$. We now focus on the
static limit surface (SLS). On the SLS, the asymptotic
time-translational Killing vector becomes null which is
mathematically written down as
\begin{equation}
g_{tt}=\rho^{2}-2M_{b}r=0.
\end{equation}
The real positive solutions of the above equation give radial
coordinates of the ergosphere.
\begin{figure}[H]
\centering
\begin{subfigure}{.3\textwidth}
\centering
\includegraphics[width=.95\linewidth]{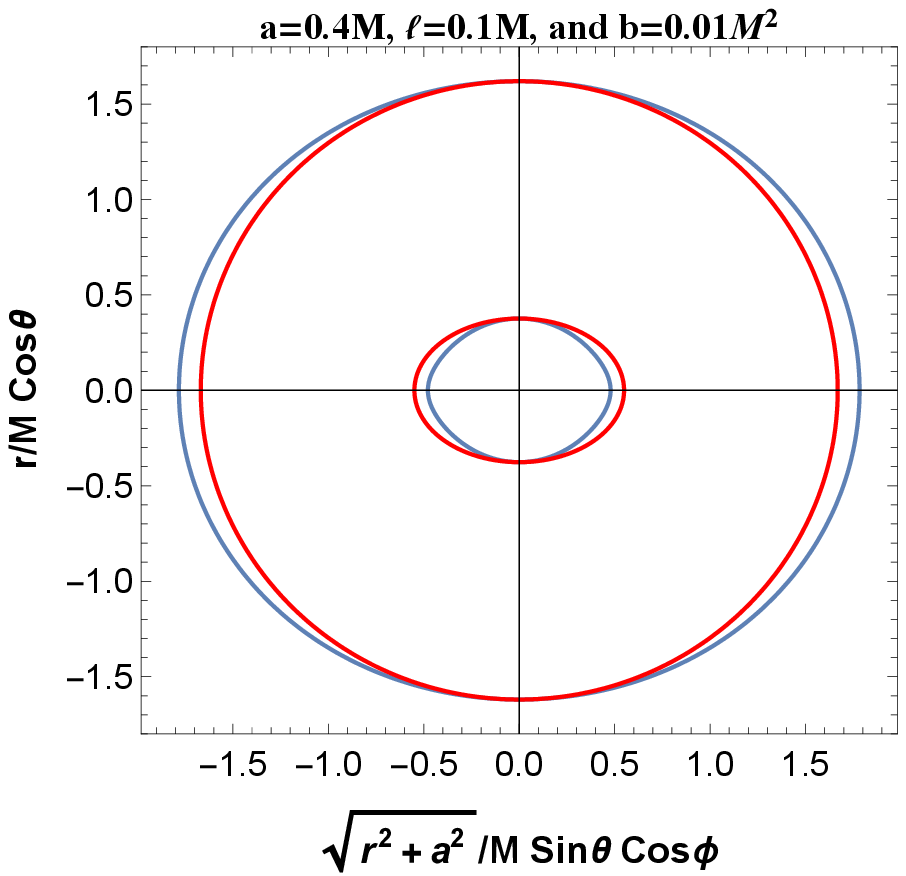}
\end{subfigure}%
\begin{subfigure}{.3\textwidth}
\centering
\includegraphics[width=.95\linewidth]{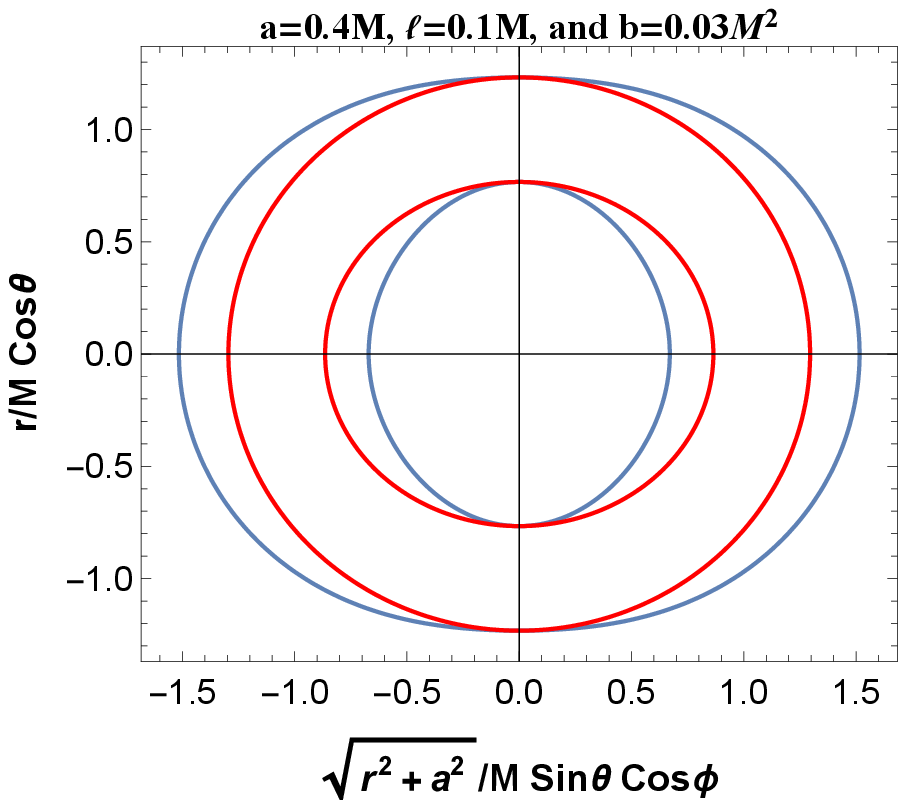}
\end{subfigure}%
\begin{subfigure}{.3\textwidth}
\centering
\includegraphics[width=.95\linewidth]{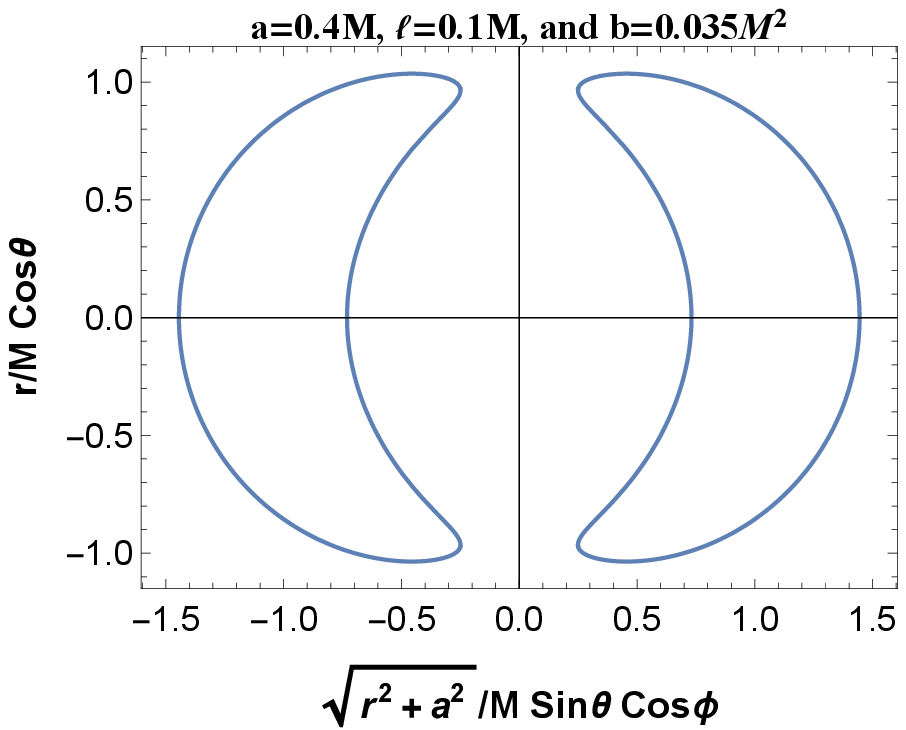}
\end{subfigure}
\par\smallskip
\begin{subfigure}{.3\textwidth}
\centering
\includegraphics[width=.95\linewidth]{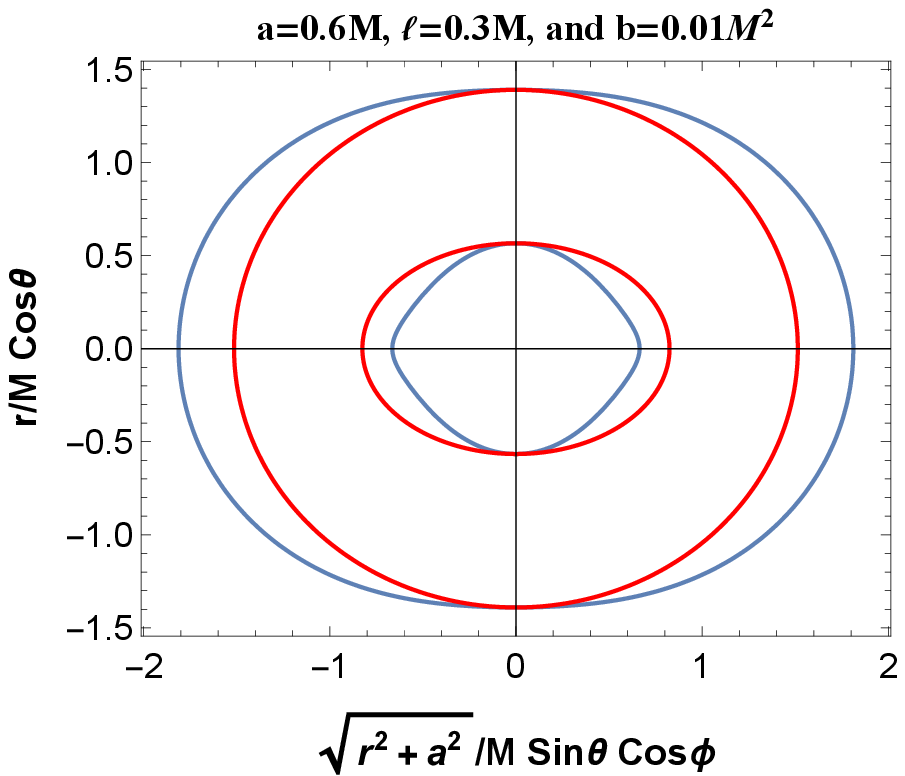}
\end{subfigure}%
\begin{subfigure}{.3\textwidth}
\centering
\includegraphics[width=.95\linewidth]{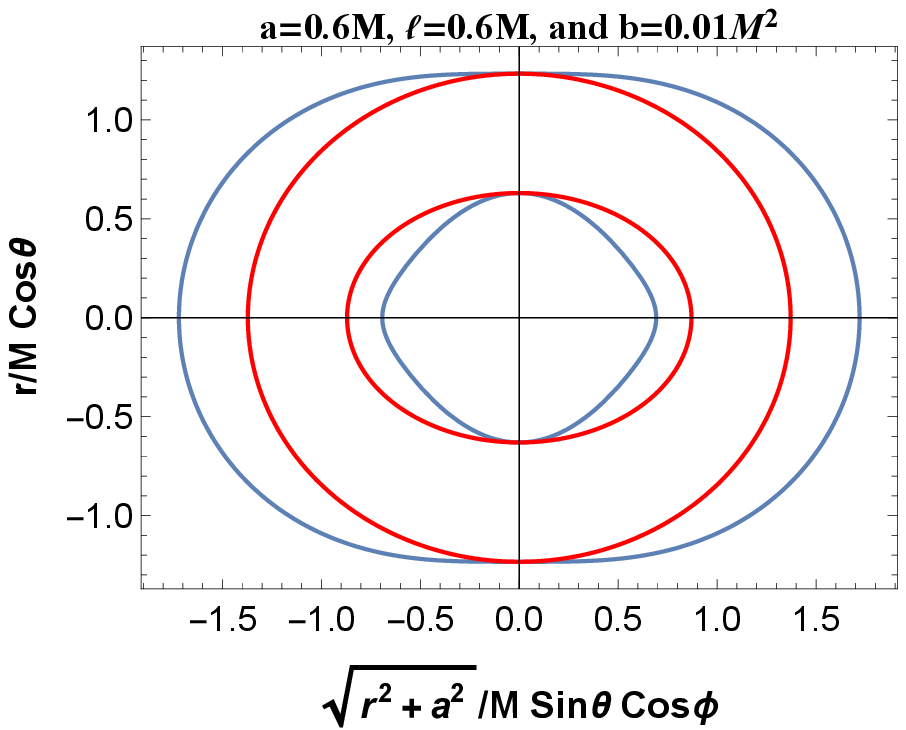}
\end{subfigure}%
\begin{subfigure}{.3\textwidth}
\centering
\includegraphics[width=.95\linewidth]{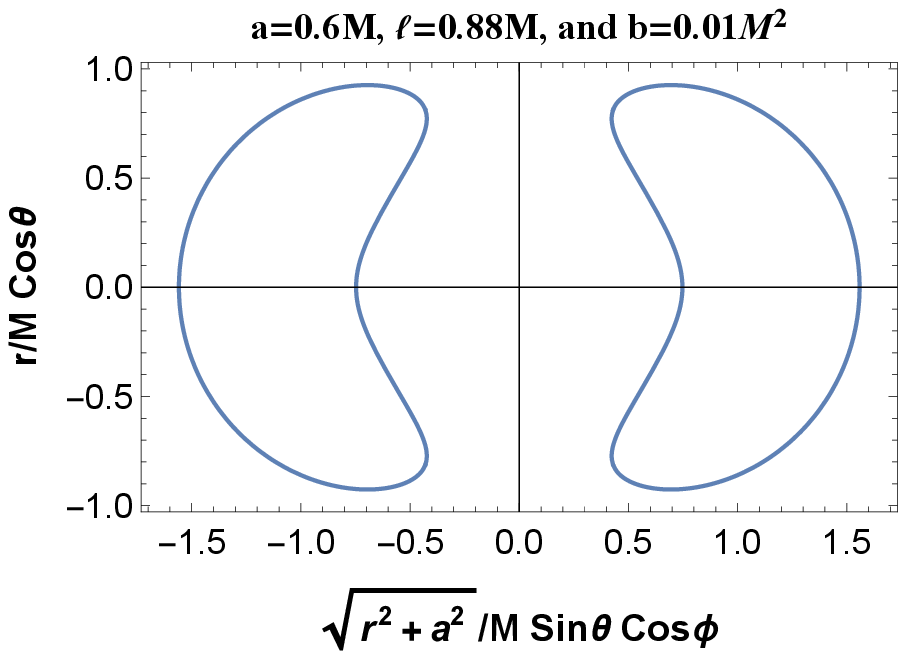}
\end{subfigure}
\caption{The cross-section of the event horizon (outer red line),
SLS (outer blue dotted line), and ergoregion of non-commutative
Kerr-like black holes.} \label{fig:ergo}
\end{figure}
The ergosphere, which lies between SLS and the event horizon, is
depicted above. As we have learned in the paper \cite{PENROSE},
energy can be extracted from the ergosphere in this situation
indeed. It reveals that the shape and size of the ergosphere
depend on
the rotational parameter $a$, the non-commutative parameter $b$, and the
parameter $\ell$. The size of the ergosphere increases with the
increase in the value of parameters $b$ and $\ell$.
\section{Superradiance scattering of the scalar field off
rotating Simpson-Visser black hole augmented with non-commutative
correction} To study the superradiance scattering of a scalar
field $\Phi$ off a non-commutative Simpson-Visser black hole field
we consider the Klein-Gordon equation corresponding to this curved
spacetime. It is given by
\begin{eqnarray}
\left(\bigtriangledown_{\alpha}\bigtriangledown^{\alpha}+\mu^{2}\right)\Phi(t,r,\theta,\phi)
= \left[\frac{-1}{\sqrt{-g}}\partial_{\sigma}\left(g^{\sigma
\tau}\sqrt{-g}\partial_{\tau}\right)+\mu^{2}\right]\Phi(t,r,\theta,\phi)&=&0,
\label{KG}
\end{eqnarray}
which is a second-order coupled differential equation. Here $\mu$
represents the mass of the scalar field $\Phi$. To separate this
coupled differential equation into radial and angular parts we now use the
standard separation of variables method for the equation
Eq.(\ref{KG}). The following ansatz in the Boyer-Lindquist
coordinates $(t, r, \theta, \phi)$
\begin{eqnarray}
\Phi(t, r, \theta, \phi)=F_{\omega j m}(r) \Theta(\theta) e^{-i
\omega t} e^{i m \phi}, \quad j \geq 0, \quad-j \leq m \leq j,
\quad \omega>0, \label{PHI}
\end{eqnarray}
is found to be useful to get the required decoupled equations.
Here $R_{\omega j m}(r)$ represents the radial function and
$\Theta(\theta)$ refers to the oblate spheroidal wave function.
The symbols $j$, $m$, and $\omega$, respectively, stand for the
angular eigenfunction, angular quantum number, and the positive
frequency of the field under investigation as viewed by a far away
observer. The use of the suitable ansatz (\ref{PHI}) enables us
to separate the differential equation (\ref{KG}) into the
following
two ordinary second-order differential equations. The radial part
of the equation reads
\begin{eqnarray}
&&\frac{d}{d r}(\Delta \frac{d F_{\omega j m}(r)}{d
r})+(\frac{((r^2+a^{2}+\ell^2) \omega-am
)^{2}}{\Delta})F_{\omega l m}(r)
\nonumber \\
&&-(\mu^{2} r^2+\mu^{2}\ell^2+j(j+1)+a^{2} \omega^{2}-2 m \omega
a) F_{\omega l m}(r)=0, \label{RE}
\end{eqnarray}
and the angular part of this is
\begin{eqnarray}
&&\sin \theta \frac{d}{d \theta}\left(\sin \theta \frac{d
\Theta_{\omega j m}(\theta)}{d \theta}\right)+\left(j(j+1) \sin
^{2} \theta-\left(\left(a \omega \sin ^{2}
\theta-m\right)^{2}\right)\right)\Theta_{\omega j m}(\theta)\nonumber \\
&& + a^{2} \mu^{2} \sin ^{2} \theta \cos ^{2} \theta~
\Theta_{\omega j m}(\theta)=0.
\end{eqnarray}
We may have a general solution of the radial equation (\ref{RE})
from the knowledge gained from the earlier investigation
\cite{BEZERRA, KRANIOTIS}, which we do not require here since we
are intended in studying the scattering of the field $\Phi$
keeping in view the superradiance phenomena with prime importance. We will use the conventional asymptotic matching procedure which has
been found useful for this type of study \cite{STRO1, STRO2, TEUK, PAGE, RAN}.
The rationale adopted in the important contributions
\cite{STRO1, STRO2, TEUK, PAGE, MK, RAN} led us to reach the required
result without going through the general solution of
the equation (\ref{RE}). The radial part of the equation (\ref{RE})
is taken into account to find an asymptotic solution. We invite
Regge-Wheeler-like coordinate $r_{*}$ which is defined below.
\begin{eqnarray}
r_{*} \equiv \int d r \frac{r^2+a^{2}+\ell^2}{\Delta},
\quad\left(r_{*} \rightarrow-\infty \quad \text{at event horizon},
\quad r_{*} \rightarrow \infty \quad \text{at infinity} \right).
\end{eqnarray}
Our requirement demands this very choice of co-ordinate as it is
beneficial to deal with the radial equation in this situation.

A new radial function $\mathcal{F}_{\omega j
m}\left(r_{*}\right)=\sqrt{r^2+a^{2}+\ell^2} F_{\omega j m}(r)$ is
introduced at this stage and after going through a few steps of
algebra we obtain the following radial equation in the desired shape
\begin{equation}
\frac{d^{2} \mathcal{F}_{\omega l m}\left(r_{*}\right)}{d
r_{*}^{2}}+V_{\omega j m}(r) \mathcal{F}_{\omega j
m}\left(r_{*}\right)=0. \label{RE1}
\end{equation}
Note that an effective potential comes into the picture which has
a crucial role in the scattering of the scalar field $\Phi$. The
effective potential reads
\begin{eqnarray}
V_{\omega j m}(r)&=&\left(\omega-\frac{ma}{r^2+a^{2}+\ell^2}\right)^{2}-\frac{\Delta}{\left(r^2
+a^{2}+\ell^2\right)^{2}}\left[\right. j(j+1)+a^{2}
\omega^{2}-2 m a \omega+\mu^{2} r^2
\\\nonumber
&&\left.+\mu^{2}\ell^2+\sqrt{r^2+a^{2}+\ell^2}\frac{d}{dr}\left(\frac{r\Delta
}{\left(r^2+a^{2}+\ell^2\right)^{\frac{3}{2}}}\right)\right],
\label{POT}
\end{eqnarray}
It turns out that this problem is comparable to the scattering of the scalar
field $\Phi$ under the effective potential
(\ref{POT}). Let us now see the asymptotic behavior of the
scattering potential at the event horizon and spatial infinity. We find that
the potential in the asymptotic limit at the event horizon acquires the
following simplified form
\begin{eqnarray}
\lim _{r \rightarrow r_{eh}} V_{\omega j m}(r)=\left(\omega-m
\Omega_{eh}\right)^{2} \equiv k_{e h}^{2},
\end{eqnarray}
and after a few steps of algebra, the potential turns into the
following at spatial infinity.
\begin{equation}
\lim _{r \rightarrow \infty} V_{\omega j m}(r)=\omega^{2}\lim_{r
\rightarrow \infty}\frac{\mu^{2}r^2\Delta}{(r^2+
a^{2}+\ell^2)^{2}} = \omega^{2}-\mu^{2}\equiv k^2_{\infty}.
\end{equation}
Here $\Omega_{eh}=\frac{a}{r_{eh}^2+\ell^2+a^2}$. Although the
potential shows constant behavior at the two extremal points
namely at the event horizon and at spatial infinity, the numerical
values of the constants are different indeed at the two extremal
points.

After scrutinizing the behavior of the potential at the two
extreme points, we now make the way to explore the asymptotic behavior of the radial solution. After a little algebra, we find
that in the asymptotic limit the radial equation (\ref{RE1}) has
the solution
\begin{equation}\label{AS}
F_{\omega j m}(r) \rightarrow\left\{\begin{array}{cl}
\frac{\mathcal{E}_{i n}^{eh} e^{-i k_{eh} r_{*}}}{\sqrt{r_{e h}^2
+a^{2}+\ell^2}} & \text { for } r \rightarrow r_{e h} \\
\mathcal{E}_{i n}^{\infty} \frac{e^{-i k_{\infty}
r_{*}}}{r}+\mathcal{E}_{r e f}^{\infty} \frac{e^{i k_{\infty}
r_{*}}}{r} & \text { for } r \rightarrow \infty,
\end{array}\right\}
\end{equation}
where $\mathcal{D}_{in}^{eh}$ is representing the amplitude of the
incoming scalar wave at event horizon($r_{eh}$) and
$\mathcal{D}_{in}^{\infty}$ stands for the amplitude of
that incoming scalar wave at infinity $(\infty)$. At spatial infinity,
the amplitude of the reflected part of the scalar wave is designated by
$\mathcal{D}_{ref}^{\infty}$. We will now be able to compute the
Wronskian which in turn will lead us to calculate its
limiting values for the region adjacent to the event horizon and
at infinity. The Wronskian for the event horizon has the
expression
\begin{equation}
W_{eh}=\left(F_{\omega j m}^{e h} \frac{d F_{\omega j m}^{* e
h}}{d r_{*}}-F_{\omega j m}^{* eh} \frac{d F_{\omega j m}^{eh}}{d
r_{*}}\right),
\end{equation}
and the Wronskian at infinity is given by the expression
\begin{equation}
W_{\infty}=\left(F_{\omega j m}^{\infty} \frac{d F_{\omega j m}^{*
\infty}}{d r_{*}}-F_{\omega j m}^{* \infty} \frac{d F_{\omega j
m}^{\infty}}{d r_{*}}\right).
\end{equation}
Note that the solutions are linearly independent. Therefore, the information concerning the standard theory of
ordinary differential equations leads us to draw an inference that
the Wronskian associated with the solutions will be independent of
$r^*$. On that account, the Wronskian evaluated at the horizon is
compatible to equate with the Wronskian evaluated at infinity. It
is in fact associated with the flux conservation of the process
\cite{REVIEW} in the physical sense. A fascinating relation
between the amplitudes of incoming and reflected waves at
different regions of interest results from equating the Wronskian.
The said relation reads
\begin{equation}
\left|\mathcal{E}_{r e f}^{\infty}\right|^{2}=\left|\mathcal{E}_{i
n}^{\infty}\right|^{2}-\frac{k_{e
h}}{k_{\infty}}\left|\mathcal{E}_{i n}^{e h}\right|^{2}.
\label{AMP}
\end{equation}
Note that if $\frac{k_{e h}}{k_{\infty}} <0$, i.e.
$\omega < m\Omega_{e h}$, the scalar wave will be superradiantly
amplified, since the relation $\left|\mathcal{E}_{r e
f}^{\infty}\right|^{2}>\left|\mathcal{E}_{i
n}^{\infty}\right|^{2}$ precisely holds in this situation.
\subsection{Amplification factor $Z_{jm}$ for superradiance}
The radial equation (\ref{RE}) which we have obtained can be express
in the following form after a few steps of algebra
\begin{eqnarray}\nonumber
&&\Delta^{2} \frac{d^{2} F_{\omega j m}(r)}{d r^{2}}+\Delta
\frac{d \Delta}{d r} \cdot \frac{d F_{\omega j m}(r)}{d r}\\
&&+\left(\left(\left(r^2+a^{2}+\ell^2\right)
\omega-a m\right)^{2}-\Delta\left(\mu^{2}
r^2+\mu^{2}\ell^2+j(j+1)+a^{2} \omega^{2}-2 m a \omega\right)\right)
F_{\omega j m}(r)=0. \label{RE2}
\end{eqnarray}
Our next step is to find out the solution for the near and the
far region and in turn make an attempt to have a single solution
by matching the solution for near-region at infinity with the solution for the far-region at its initial point such that this
particular solution be amenable in the vicinity of the cardinal
region. At this stage, it is convenient to make use of a new
variable $y$ which is defined by
$u=\frac{r-r_{eh}}{r_{eh}-r_{ch}}$. In terms of $y$ the equation
(\ref{RE2}) can be written down as
\begin{eqnarray}
&&u^{2}(u+1)^{2} \frac{\mathrm{d}^{2} F_{\omega j
m}(u)}{\mathrm{d} u^{2}}+u(u+1)(2 u+1) \frac{\mathrm{d} F_{\omega
j m}(y)}{\mathrm{d} u} \\\nonumber &&+\left(\mathcal{Q}^2
u^{4}+B^{2}-j(j+1) u(u+1)-\frac{\mu^{2}
\mathcal{Q}^{2}}{\omega^{2}} u^{3}(u+1)-\mu^{2} r_{e h}^{2}
u(u+1)-\frac{2 \mu^{2} r_{e h} \mathcal{Q}}{\omega} u^{2}(u+1)
\right)F_{\omega j m}(y)=0, \label{AAPP}
\end{eqnarray}
with the approximation $a \omega \ll 1$. In equation (\ref{AAPP}, the
$B$ and $B\mathcal{Q}$ have the following
expressions:
\begin{eqnarray}
B&=&\frac{(\omega-m \Omega_{eh}}{r_{e h}-r_{c h}} r_{e h}^{2} \nonumber \\
\mathcal{Q}&=&\left(r_{e h}-r_{c h}\right) \omega.
\end{eqnarray}
Let us now focus on the near-region solution where we have
$\mathcal{Q} y \ll 1$ and $\mu^{2} r_{e h}^{2} \ll 1$. Therefore,
for the near-region the above equation gets a simplified form:
\begin{eqnarray}
u^{2}(u+1)^{2} \frac{\mathrm{d}^{2} F_{\omega j m}(y)}{\mathrm{d}
u^{2}}+u(u+1)(2 u+1) \frac{\mathrm{d} F_{\omega j
m}(u)}{\mathrm{d} u}+\left(B^{2}-j(j+1)u(u+1)\right) F_{\omega j m}(r)=0.
\end{eqnarray}
Since the Compton wavelength of the boson participating in the
scattering process is much smaller than the size of the black hole the
approximation $\left(\mu^{2} r_{e h}^{2} \ll 1\right)$ is regarded in
the process. So, the general solution of the above equation in terms
of associated Legendre function of the first kind $P_{\lambda}^{\nu}(y)$
can be expressed as
\begin{eqnarray}
F_{\omega j m}(u)=d
P^{2iB}_{\frac{\sqrt{1+4j(j+1)}-1}{2}}(1+2u).
\end{eqnarray}
Now, we use the relation
\begin{equation}
P_{\lambda}^{\nu}(z)=\frac{1}{\Gamma(1-\nu)}\left(\frac{1+z}{1-z}\right)^{\nu
/ 2}{ }_{2} F_{1}\left(-\lambda, \lambda+1 ; 1-\nu ;
\frac{1-z}{2}\right),
\end{equation}
that facilitates us to express $F_{\omega j m}(y)$ in terms of
the ordinary hypergeometric functions ${ }_{2} F_{1}(a, b ; c ;
z)$ :
\begin{equation}
F_{\omega j m}(u)=d\left(\frac{u}{u+1}\right)^{-i B}{
}_{2} F_{1}\left(\frac{1-\sqrt{1+4j(j+1)}}{2},
\frac{1+\sqrt{1+4 j(j+1)}}{2} ; 1-2 i B
;-u\right).\label{NEAR}
\end{equation}
We have mentioned already that we are intended to design a single solution
using the matching condition at the desired (cardinal) position where the
two solutions mingle with each other. Therefore, we need to have
the behavior of the above expression (\ref{NEAR}) for large $u$. For large u, i.e. $(u \to\infty$) the Eq.(\ref{NEAR}) turns into
\begin{eqnarray}
F_{\text {near-large } u} \sim d &&\left(\frac{\Gamma(\sqrt{1+4 j(j+1)})
\Gamma(1-2 i B)}{\Gamma\left(\frac{1+\sqrt{1+4 j(j+1)}}{2}
-2 i B\right) \Gamma\left(\frac{1+\sqrt{1+4 j(j+1)}}{2}\right)}
u^{\frac{\sqrt{1+4 j(j+1)}-1}{2}}+\right.\\
&& \frac{\Gamma(-\sqrt{1+4 j(j+1)}) \Gamma(1-2
i B)}{\Gamma\left(\frac{1-\sqrt{1+4j(j+1)}}{2}\right)
\Gamma\left(\frac{1-\sqrt{1+4j(j+1)}}{2}-2 iB\right)}
u^{\left.-\frac{\sqrt{1+4 j(j+1)}+1}{2}\right)} .
\label{NF}
\end{eqnarray}
Let us look at the solution for the far region. To match the
solution of far and near regions at the cardinal region the
behavior of solution (\ref{RE2}) with the approximations $u+1
\approx u$ and $\mu^{2} r_{e h}^{2} \ll 1$ is to be computed. We
may drop all the terms except those which describe the free motion
with momentum $j$ and after the dropping out of all these terms
the equation (\ref{RE2}) reduces to
\begin{equation}
\frac{\mathrm{d}^{2} F_{\omega j m}(u)}{\mathrm{d}
u^{2}}+\frac{2}{u} \frac{\mathrm{d} F_{\omega j m}(u)}{\mathrm{d}
u}+\left(k_{l}^{2}-\frac{j(j+1)}{u^{2}}\right) F_{\omega j
m}(u)=0, \label{FAR}
\end{equation}
where $k_{l} \equiv \frac{\mathcal{Q}}{\omega}
\sqrt{\omega^{2}-\mu^{2}}$. Note that the equation (\ref{FAR}) can
be solved exactly and the most general solution of the equation
(\ref{FAR}) is
\begin{eqnarray}
F_{\omega j m, \text { far }}=e^{-i k y}(f_{1}
u^{\frac{\sqrt{1+4 j(j+1)}-1}{2}} M(\frac{1+\sqrt{1+4 j(j+1)}}{2},
1+\sqrt{1+4j(j+1)}, 2 i k_{l} u)+ \\
f_{2} u^{-\frac{\sqrt{1+4 j(j+1)}+1}{2}}
M(\frac{1-\sqrt{1+4 j(j+1)}}{2}, 1-\sqrt{1+4j(j+1)}, 2 i k_{l} u)), \nonumber \label{FARR}
\end{eqnarray}
where $M(a; b; y)$ is representing the confluent hypergeometric
Kummer function of the first kind. In order to equate the solution
(\ref{FARR}) with the far-region limiting solution of (\ref{NEAR})
standing in Eq. (\ref{NF}), we look for the small $y$ behavior of
the solution (\ref{FARR}). See that the equation (\ref{FARR}) turns into
\begin{equation}
F_{\omega j m, \text { far-small } \mathrm{u}} \sim u^{-\frac{1+\sqrt{1+4 j(j+1)}}{2}}, \label{FN}
\end{equation}
for small u, i.e. $(u \to 0)$. The solution (\ref{NF}) and
(\ref{FN}) are susceptible to match, since these two have a common
mingling region. The matching of the asymptotic solutions
(\ref{NF}) and (\ref{FN}) facilitates us to compute the scalar
wave flux at infinity which reads
\begin{eqnarray}
f_{1}=& d \frac{\Gamma(\sqrt{1+4 j(j+1)}) \Gamma(1-2
i B)}{\Gamma\left(\frac{1+\sqrt{1+4j(j+1)}}{2} -2 i B\right)
\Gamma\left(\frac{1+\sqrt{1+4 j(j+1)}}{2}\right)},
\\\nonumber f_{2}=& d \frac{\Gamma(-\sqrt{1+4 j(j+1)})
\Gamma(1-2 iB)}{\Gamma\left(\frac{1-\sqrt{1+4 j(j+1)}}{2}-2
i B\right) \Gamma\left(\frac{1-\sqrt{1+4j(j+1)}}{2}\right)}. \label{DD}
\end{eqnarray}
If we expand equation (\ref{FARR}) around infinity it results
\begin{eqnarray}
f_{1} \frac{\Gamma(1+\sqrt{1+4 j(j+1)})}{\Gamma\left(\frac{1+\sqrt{1+4j(j+1)}}{2}\right)} k_{l}^{-\frac{1+\sqrt{1+4j(j+1)}}{2}}\left((-2 i)^{-\frac{1+\sqrt{1+4 j(j+1)}}{2}} \frac{e^{-i k_{l} u}}{u}+(2 i)^{-\frac{1+\sqrt{1+4 j(j+1)}}{2}}
\frac{e^{i k_{l} u}}{u}\right)+ \\\nonumber
f_{2} \frac{\Gamma(1-\sqrt{1+4j(j+1)})}{\Gamma\left(\frac{1-\sqrt{1+4 j(j+1)}}{2}\right)}
k_{l}^{\frac{\sqrt{1+4 j(j+1)}-1}{2}}\left((-2
i)^{\frac{\sqrt{1+4 j(j+1)}-1}{2}} \frac{e^{-i k_{l}
u}}{u}+(2 i)^{\frac{\sqrt{1+4 j(j+1)}-1}{2}} \frac{e^{i
k_{l} u}}{u}\right), \label{FARRI}
\end{eqnarray}
if the approximations $\frac{1}{u} \sim \frac{\mathcal{Q}}{\omega}
\cdot \frac{1}{r}, \quad e^{\pm i k_{l} u} \sim e^{\pm
i\sqrt{(\omega^{2}-\mu^{2})}r} $ are made functional. If a
comparison of Eq. (\ref{FARRI}) is made with the radial solution
\eqref{AS} we have
$$
F_{\infty}(r) \sim \mathcal{E}_{i n}^{\infty} \frac{e^{-i
\sqrt{\omega^{2}-\mu^{2}}
r^{*}}}{r}+\mathcal{E}_{r e f}^{\infty} \frac{e^{i
\sqrt{\omega^{2}-\mu^{2}} r^{*}}}{r}, \quad
\text { for } \quad r \rightarrow \infty,
$$
where
$$
\begin{array}{c}
\mathcal{E}_{i n}^{\infty}=\frac{\mathcal{Q}}{\omega}\left(f_{1}(-2 i)^{-\frac{1+\sqrt{1+4j(j+1)}}{2}}
\frac{\Gamma(1+\sqrt{1+4j(j+1)})}{\Gamma\left(\frac{1+\sqrt{1+4j(j+1)}}{2}\right)}
k_{l}^{-\frac{1+\sqrt{1+4j(j+1)}}{2}}+\right. \\
\left.f_{2}(-2 i)^{\frac{\sqrt{1+4j(j+1)}-1}{2}}
\frac{\Gamma(1-\sqrt{1+4j(j+1)})}{\Gamma\left(\frac{1-\sqrt{1+4j(j+1)}}{2}\right)}
k_{l}^{\frac{\sqrt{1+4j(j+1)}-1}{2}}\right),
\end{array}
$$
and
$$
\begin{array}{l}
\mathcal{E}_{r e f}^{\infty}=\frac{\mathcal{Q}}{\omega}\left(f_{1}(2 i)^{-\frac{1+\sqrt{1+4j(j+1)}}{2}}
\frac{\Gamma(1+\sqrt{1+4j(j+1)})}{\Gamma\left(\frac{1+\sqrt{1+4j(j+1)}}{2}\right)}
k_{l}^{-\frac{1+\sqrt{1+4j(j+1)}}{2}}+\right. \\
\left.f_{2}(2 i) \frac{\sqrt{1+4j(j+1)}-1}{2}
\frac{\Gamma(1-\sqrt{1+4j(j+1)})}{\Gamma\left(\frac{1-\sqrt{1+4j(j+1)}}{2}\right)}
k_{l}^{\frac{\sqrt{1+4j(j+1)}-1}{2}}\right).
\end{array}
$$
Substituting the expressions of $f_{1}$ and $f_{2}$ from Eq.
(\ref{DD}) into the above expressions, we land onto
\begin{eqnarray}
\mathcal{E}_{in}^{\infty}&=&\frac{d(-2
i)^{-\frac{1+\sqrt{1+4j(j+1)}}{2}}}{\sqrt{(\omega^{2}-\mu^{2})}} \cdot
\frac{\Gamma(\sqrt{1+4 j(j+1)}) \Gamma(1+\sqrt{1+4j(j+1)})}{\Gamma\left(\frac{1+\sqrt{1+4j(j+1)}}{2}-2
iB\right)\left(\Gamma\left(\frac{1+\sqrt{1+4j(j+1)}}{2}\right)\right)^{2}}\times
\\\nonumber &&\Gamma(1-2 i B)
k_{l}^{\frac{1-\sqrt{1+4j(j+1)}}{2}}+\frac{d(-2
i)^{\frac{\sqrt{1+4j(j+1)}-1}{2}}}{\sqrt{(\omega^{2}-\hat{\mu}^{2})}}
\times \\\nonumber &&\frac{\Gamma(1-\sqrt{1+4j(j+1)})
\Gamma(-\sqrt{1+4j(j+1)})}{\left(\Gamma\left(\frac{1-\sqrt{1+4j(j+1)}}{2}\right)\right)^{2}
\Gamma\left(\frac{1-\sqrt{1+4j(j+1)}}{2}-2 iB\right)} \Gamma(1-2 iB)
k_{l}^{\frac{1+\sqrt{1+4j(j+1)}}{2}},
\end{eqnarray}
and
\begin{eqnarray}
\mathcal{E}_{ref}^{\infty}&=&\frac{d(2
i)^{-\frac{1+\sqrt{1+4j(j+1)}}{2}}}{\sqrt{(\omega^{2}-\mu^{2})}} \cdot
\frac{\Gamma(\sqrt{1+4 j(j+1)}) \Gamma(1+\sqrt{1+4j(j+1)})}{\Gamma\left(\frac{1+\sqrt{1+4j(j+1)}}{2}-2iB\right)\left(\Gamma\left(\frac{1+\sqrt{1+4j(j+1)}}{2}\right)\right)^{2}}\times
\\\nonumber &&\Gamma(1-2 i B)
k_{l}^{\frac{1-\sqrt{1+4j(j+1)}}{2}}+\frac{d(2
i)^{\frac{\sqrt{1+4j(j+1)}-1}{2}}}{\sqrt{(\omega^{2}-\mu^{2})}}
\times \\\nonumber &&\frac{\Gamma(1-\sqrt{1+4j(j+1)})
\Gamma(-\sqrt{1+4j(j+1)})}{\left(\Gamma\left(\frac{1-\sqrt{1+4j(j+1)}}{2}\right)\right)^{2}
\Gamma\left(\frac{1-\sqrt{1+4j(j+1)}}{2}-2 iB\right)} \Gamma(1-2 i B)
k_{l}^{\frac{1+\sqrt{1+4j(j+1)}}{2}}.
\end{eqnarray}
Ultimately, we reach the expression of the amplification factor
that results out to be
\begin{equation}
Z_{j m} \equiv \frac{\left|\mathcal{E}_{r e
f}^{\infty}\right|^{2}}{\left|\mathcal{E}_{i
n}^{\infty}\right|^{2}}-1. \label{AMPZ}
\end{equation}
Equation (\ref{AMPZ}) is a general expression of the amplification
factor obtained by making use of the asymptotic matching method. A
striking feature follows if
$\frac{\left|\mathcal{E}_{ref}^{\infty}\right|^{2}}{\left|\mathcal{E}_{in}^{\infty}\right|^{2}}$
acquires a value greater than unity. There will be a definite gain
in the amplification factor that corresponds to superradiance
phenomena. However, a negative value of the amplification factor
indicates a dropping down of the amplification that corresponds to
the non-occurrence of the superradiance phenomena. In
Fig.~(\ref{zl}), we have given a graphical presentation of the
variation $Z_{j m}$ with $M\omega$. Here the leading multi-poles
$j= 1$, and $2$ have been taken into consideration. Taking
different values of the parameter $\ell$ the plots for the
multi-poles $j= 1$, and $2$ have been displayed in Fig.~(\ref{zl})

\begin{figure}[H]
\centering
\begin{subfigure}{.58\textwidth}
\includegraphics[width=.8\linewidth]{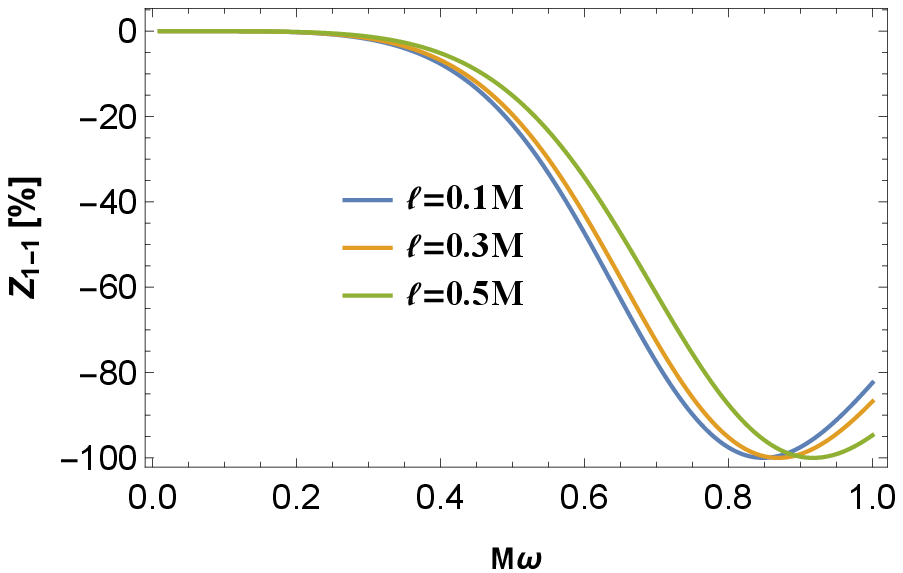}
\end{subfigure}%
\begin{subfigure}{.58\textwidth}
\includegraphics[width=.8\linewidth]{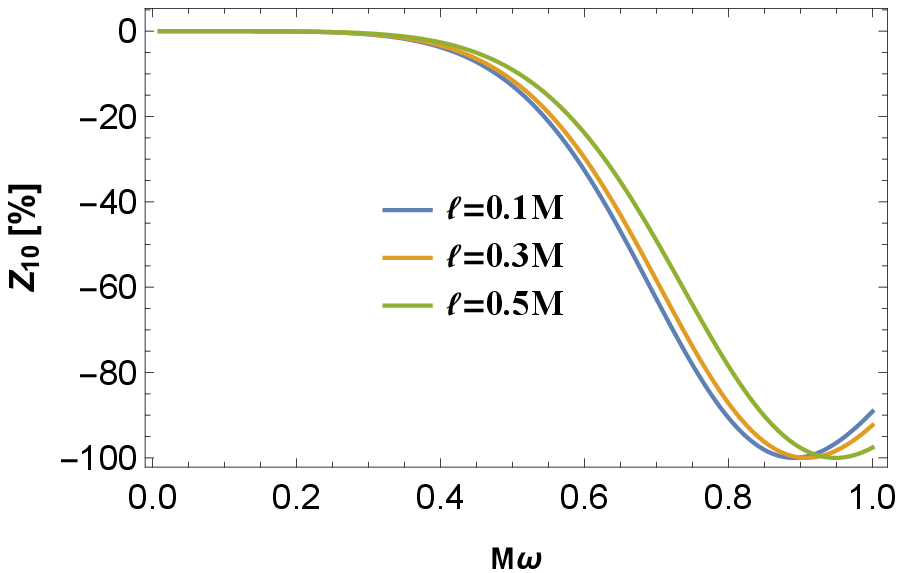}
\end{subfigure}
\caption{Variation of amplification factors with $\ell$ for
non-superradiant multi-poles with $\mu=0.1, b=0.01M^2$, and
$a=0.52M$.} \label{non-superradiant}
\end{figure}

\begin{figure}[H]
\centering
\begin{subfigure}{.58\textwidth}
\includegraphics[width=.8\linewidth]{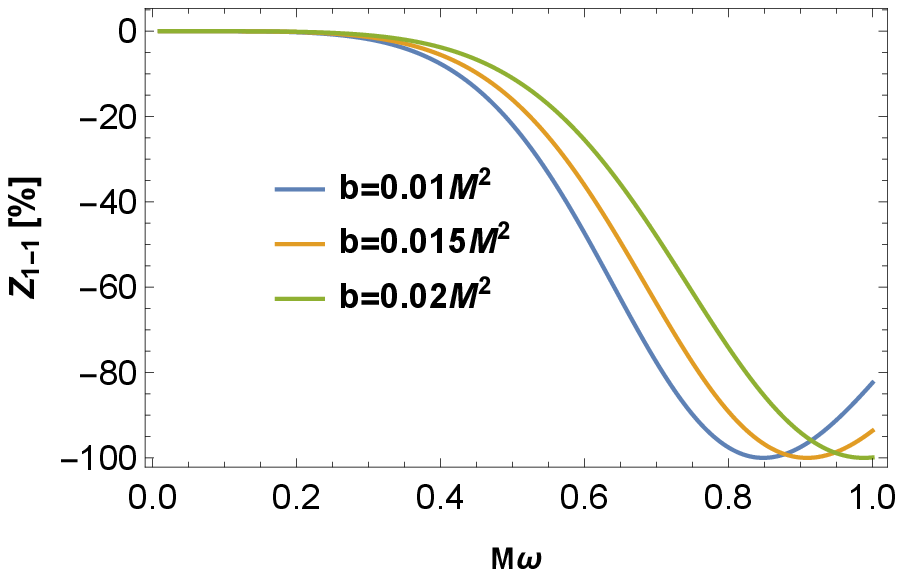}
\end{subfigure}%
\begin{subfigure}{.58\textwidth}
\includegraphics[width=.8\linewidth]{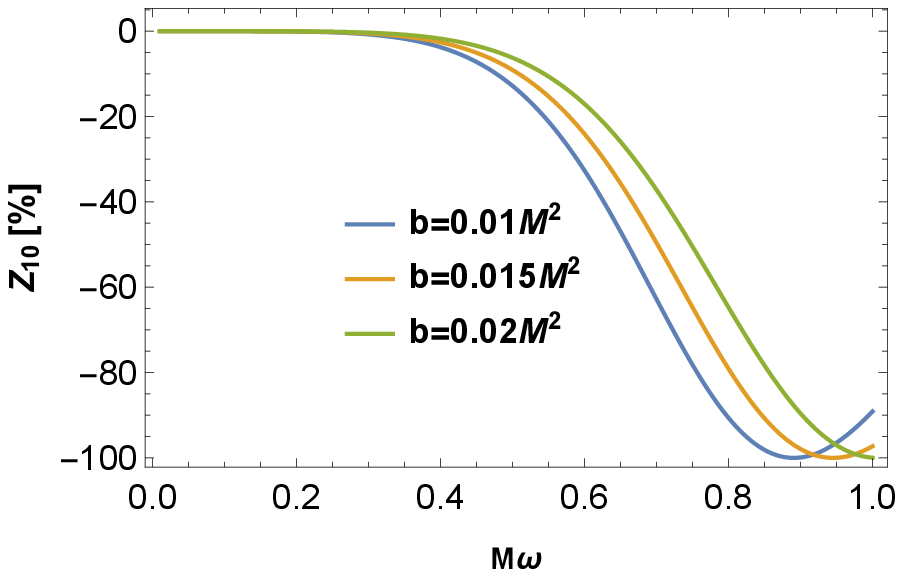}
\end{subfigure}
\caption{Variation of amplification factors with $b$ for
non-superradiant multi-poles with $\mu=0.1, \ell=0.1M$, and
$a=0.42M$.} \label{non-superradiant1}
\end{figure}

\begin{figure}[H]
\centering
\begin{subfigure}{.58\textwidth}
\includegraphics[width=.8\linewidth]{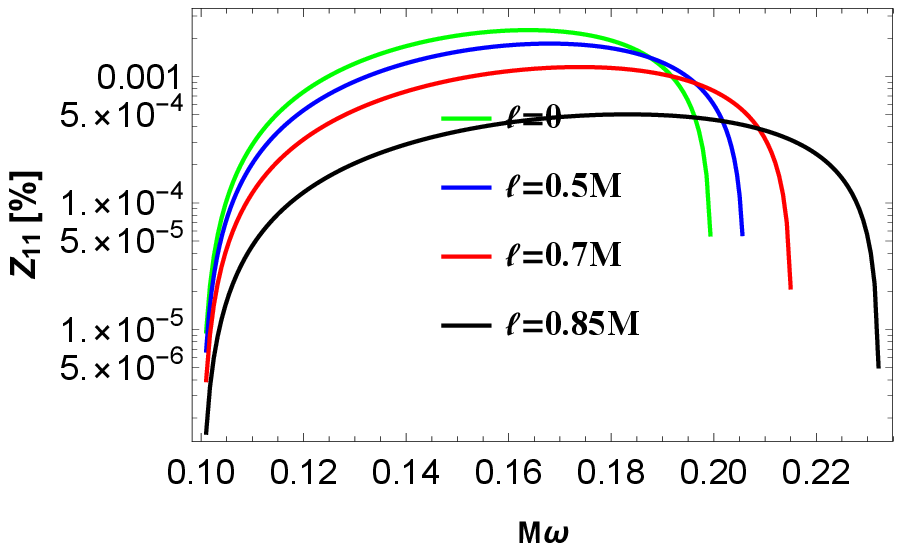}
\end{subfigure}%
\begin{subfigure}{.58\textwidth}
\includegraphics[width=.8\linewidth]{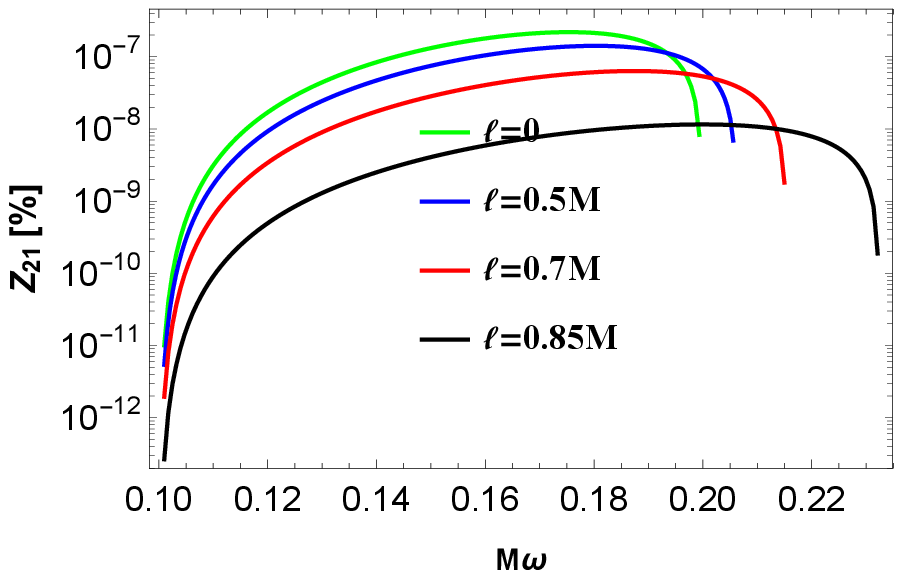}
\end{subfigure}
\begin{subfigure}{.58\textwidth}
\centering
\includegraphics[width=.8\linewidth]{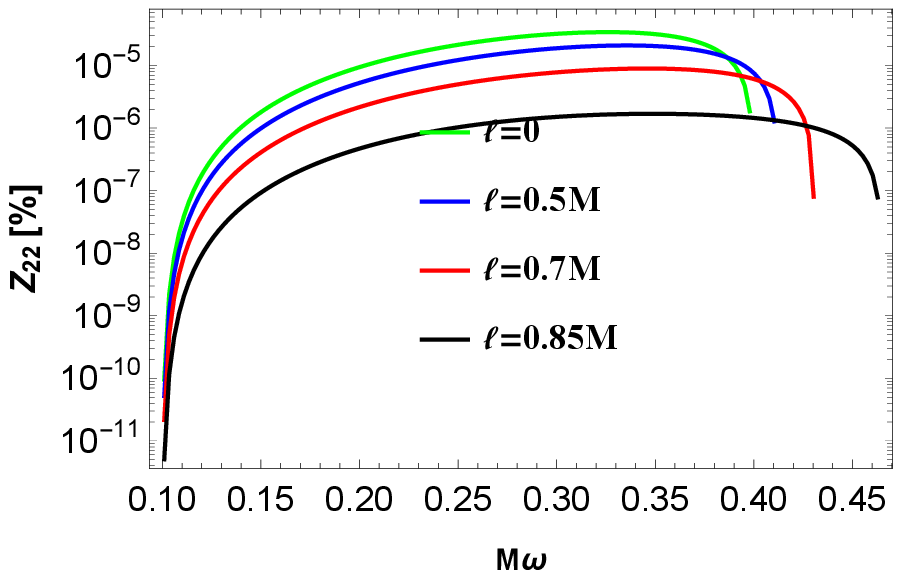}
\end{subfigure}
\caption{Variation of amplification factors with $\ell$ for
various multi-poles with $\mu=0.1, b=0.01M^2$, and $a=0.52M$.}
\label{zl}
\end{figure}

\begin{figure}[H]
\centering
\begin{subfigure}{.58\textwidth}
\includegraphics[width=.8\linewidth]{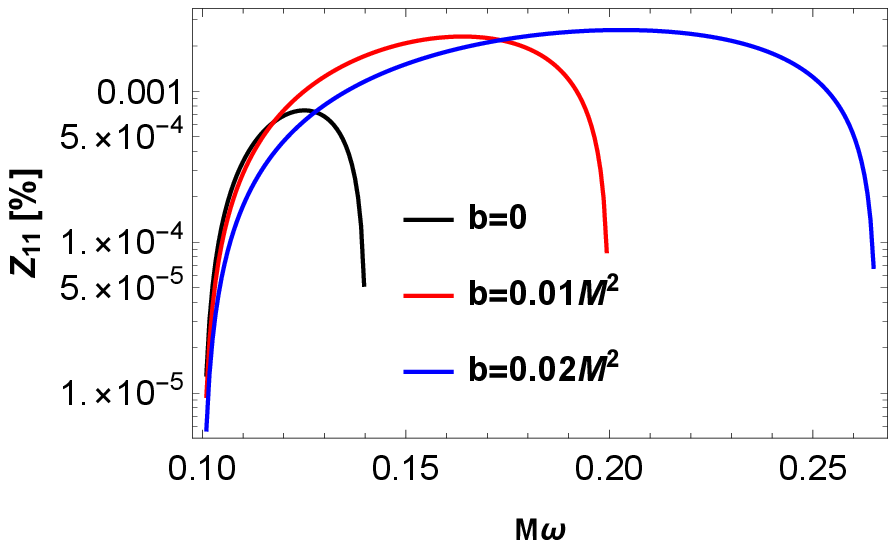}
\end{subfigure}%
\begin{subfigure}{.58\textwidth}
\includegraphics[width=.8\linewidth]{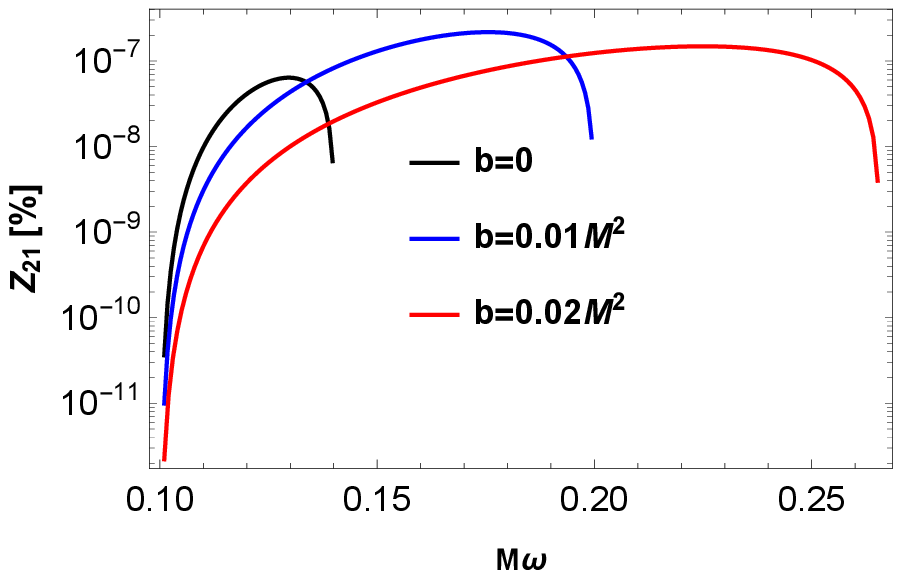}
\end{subfigure}
\begin{subfigure}{.58\textwidth}
\centering
\includegraphics[width=.8\linewidth]{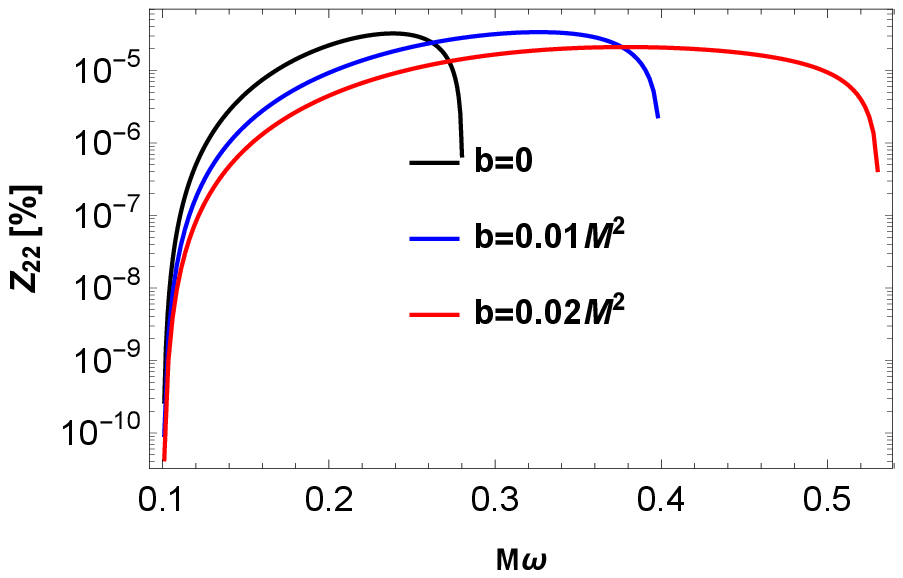}
\end{subfigure}
\caption{Variation of amplification factors with $b$ for various
multi-poles with $\mu=0.1, \ell=0.1M$, and $a=0.52M$.} \label{zb}
\end{figure}

\begin{figure}[H]
\centering
\begin{subfigure}{.58\textwidth}
\includegraphics[width=.8\linewidth]{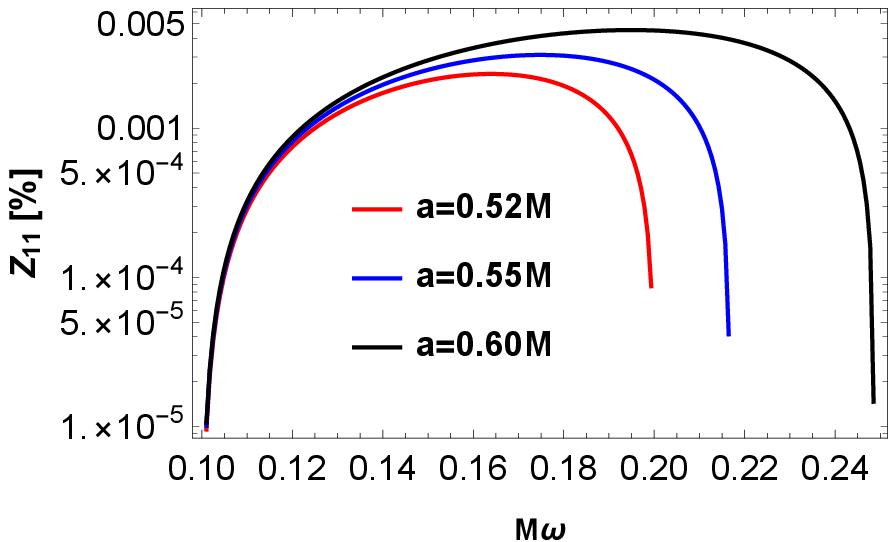}
\end{subfigure}%
\begin{subfigure}{.58\textwidth}
\includegraphics[width=.8\linewidth]{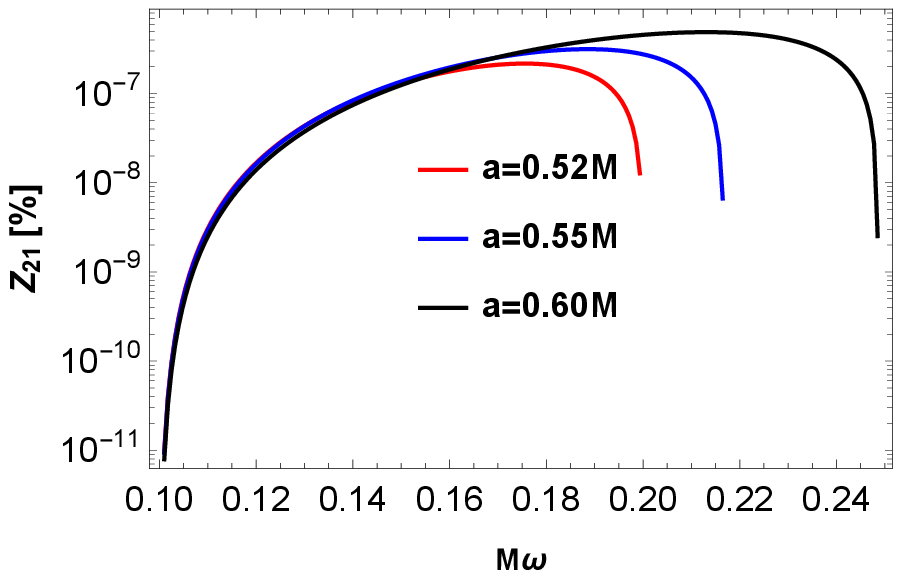}
\end{subfigure}
\begin{subfigure}{.58\textwidth}
\centering
\includegraphics[width=.8\linewidth]{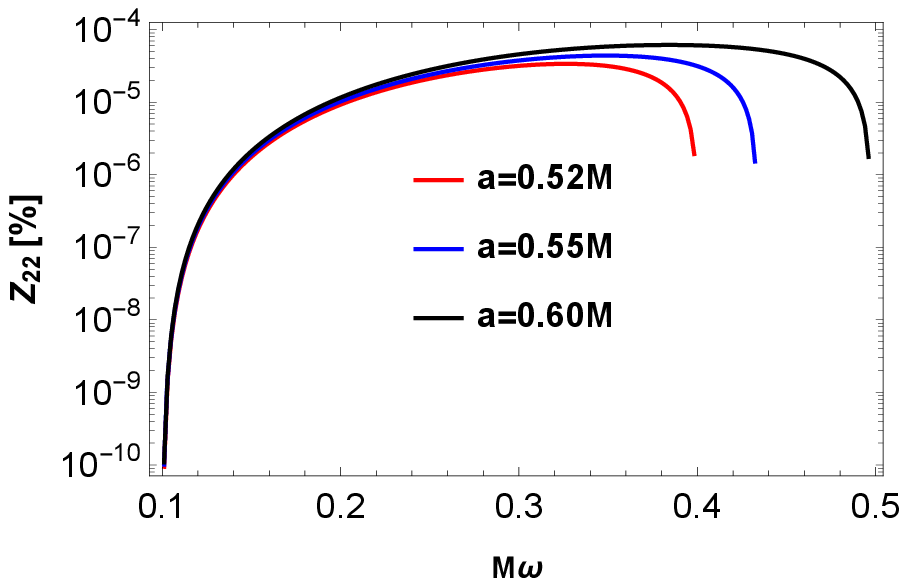}
\end{subfigure}
\caption{Variation of amplification factors with $a$ for various
multi-poles with $\mu=0.1, b=0.01M^2$, and $\ell=0.1M$.}
\label{za}
\end{figure}

A judicious look on Fig.~(\ref{non-superradiant}) and
Fig.~(\ref{non-superradiant1}), makes it evident that
superradiance for a particular value of $j$ occurs when the
allowed azimuthal values for $J$, i.e. the values of $m$ are
restricted to $m > 0$. For negative value $m$, however, the
amplification factor takes a negative value which refers to the
nonoccurrence of superradiance. It also becomes transparent from
Fig.~(\ref{zl}) that if the numerical value of the parameter
$\ell$ increases the superradiance process gets enhanced and the
reverse is the case when the value of the parameter $\ell$ drops
down. Fig.~(\ref{za}) shows the effect of the spin parameter
$a$ on the superradiance process. Here it is clearly observed that
the superradiance process intensifies with an increase in the spin
parameter $a$. The quantum effect of gravity which is amended in
our model is characterized by non-commutative parameter $b$.
Fig.~(\ref{zb}) shows that the superradiance scenario decreases with the increase in the value of the non-commutative
parameter $b$. Therefore, amendment of the quantum gravity effect
makes the superradiance process moderate.

\subsection{Superradiant instability for
Simpson-Visser black hole in the non-commutative setting} If the
superradiant field gets trapped in an enclosed system, then the
stability of the background black hole will be disturbed as a
result of energy extraction. In this situation, the scalar field
interacts with the background black hole repeatedly and the
scattered wave amplitude increases exponentially which leads to
superradiant instability. The possible occurrence of this type of
instability was first reported by Press and Teukolsky and they
termed this phenomenon as black hole bomb \cite{PRESS, TEUK}. Let
us look at (\ref{RE}) and write it as follows
\begin{eqnarray}
\Delta \frac{d}{d r}\left(\Delta \frac{d F_{\omega j m}}{d
r}\right)+\mathcal{G} F_{\omega j m}=0, \label{MRE}
\end{eqnarray}
were $\mathcal{G}$ has the form
$$
\mathcal{G} \equiv\left(\left(r^2+a^{2}+\ell^2\right) \omega-m
a\right)^{2}+\Delta\left(2 m a
\omega-j(j+1)-\mu^{2}r^2-\mu^2\ell^2\right).
$$
for a slowly rotating black hole, i.e when $a \omega \ll 1$. So,
black hole bomb mechanism will be feasible if we have the
following solutions for the radial equation (\ref{MRE})
$$
F_{\omega j m} \sim\left\{\begin{array}{ll}
e^{-i(\omega-m \Omega_{eh}) r_{*}} & \text { as } r \rightarrow r_{e h}
\left(r_{*} \rightarrow-\infty\right) \\
\frac{e^{-\sqrt{\mu^{2}-\omega^{2}} r_{*}}}{r} & \text { as } r
\rightarrow \infty \left(r_{*} \rightarrow \infty\right)
\end{array}\right.
$$
The above solution corresponds to a physical boundary condition
that the scalar wave at the black hole horizon is purely in-going
while at spatial infinity it is decaying only exponentially,
provided the condition $\omega^{2} <\ mu^{2}$ is maintained.

Now, in terms of the new radial function
$$
\psi_{\omega j m} \equiv \sqrt{\Delta} F_{\omega j m},
$$
the radial equation (\ref{MRE}) can be expressed as
\begin{equation}
(\frac{d^2r}{d r^2}+ \omega^2-V) \psi_{\omega jm}=0. \label{REGW}
\end{equation}
with
$$
\omega^{2}-V=\frac{\mathcal{G}+\frac{1}{4} \left(-\frac{8 \sqrt{b} M \sqrt{\ell^2+r^2}}{\sqrt{\pi } r^2}
-\frac{2 r \left(M-\frac{4 \sqrt{b} M}{\sqrt{\pi } r}\right)}{\sqrt{\ell^2+r^2}}+2 r\right)^2}{\Delta^{2}},
$$
\begin{figure}[H]
\centering
\begin{subfigure}{.34\textwidth}
\includegraphics[width=.9\linewidth]{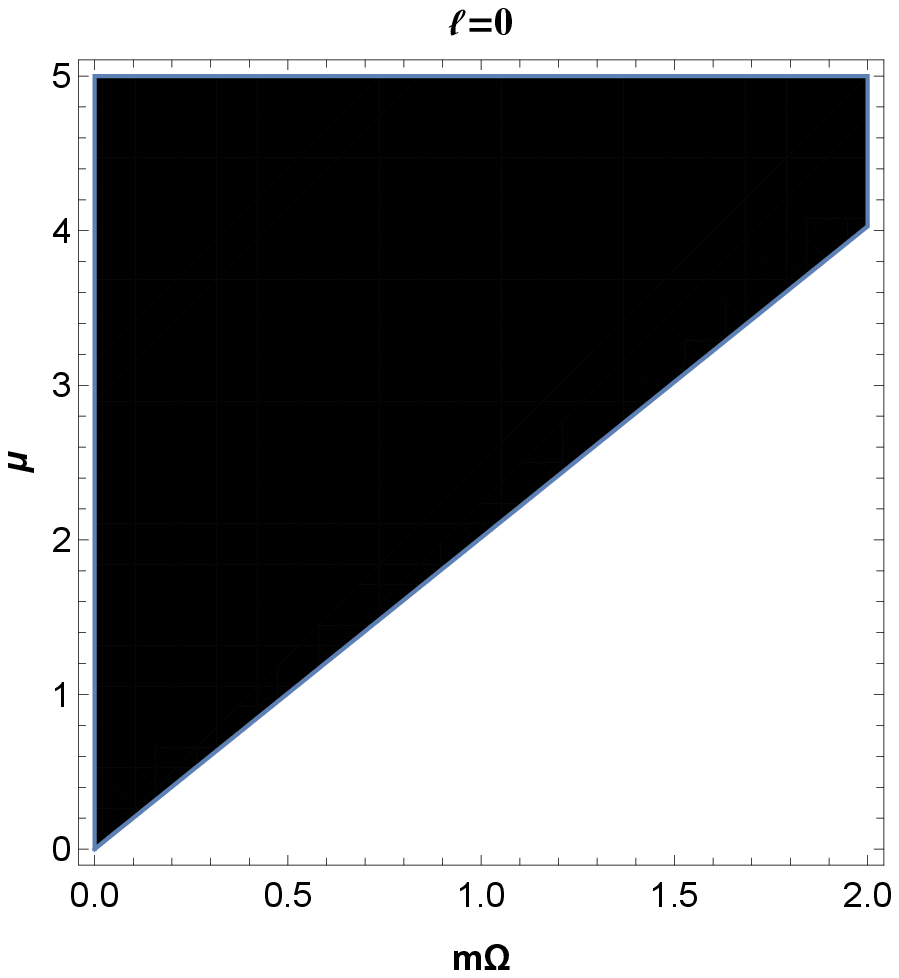}
\end{subfigure}%
\begin{subfigure}{.34\textwidth}
\includegraphics[width=.9\linewidth]{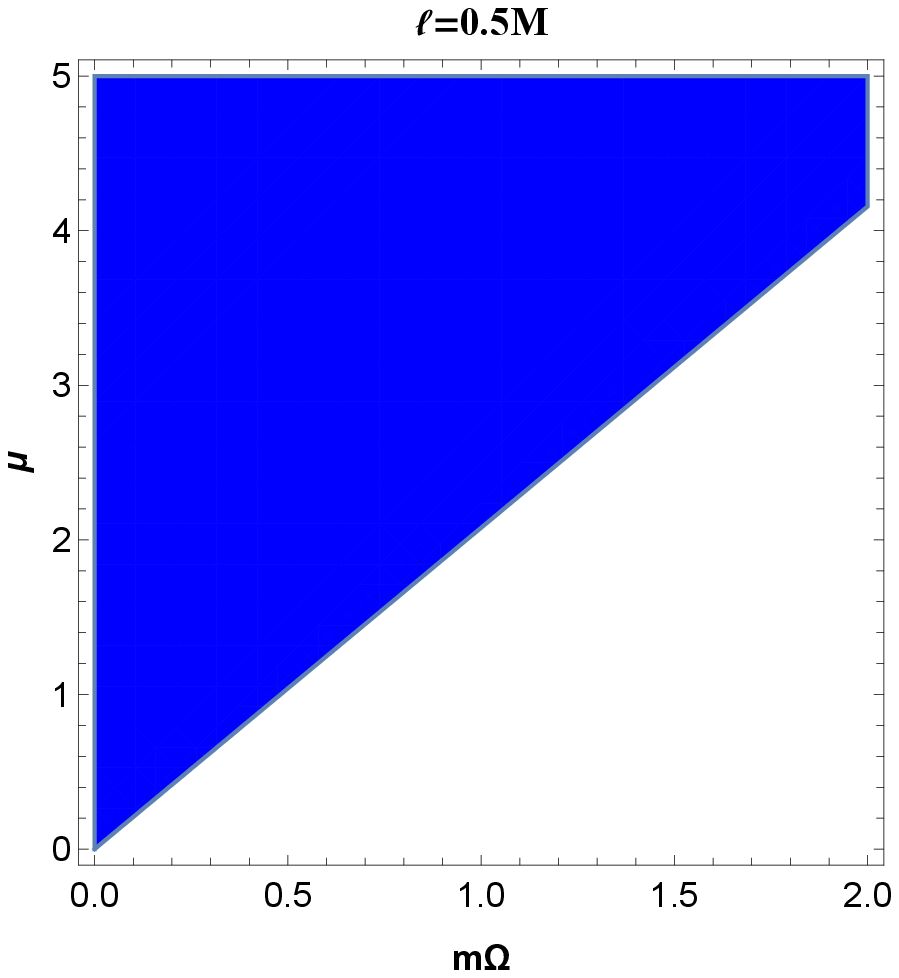}
\end{subfigure}%
\begin{subfigure}{.34\textwidth}
\includegraphics[width=.9\linewidth]{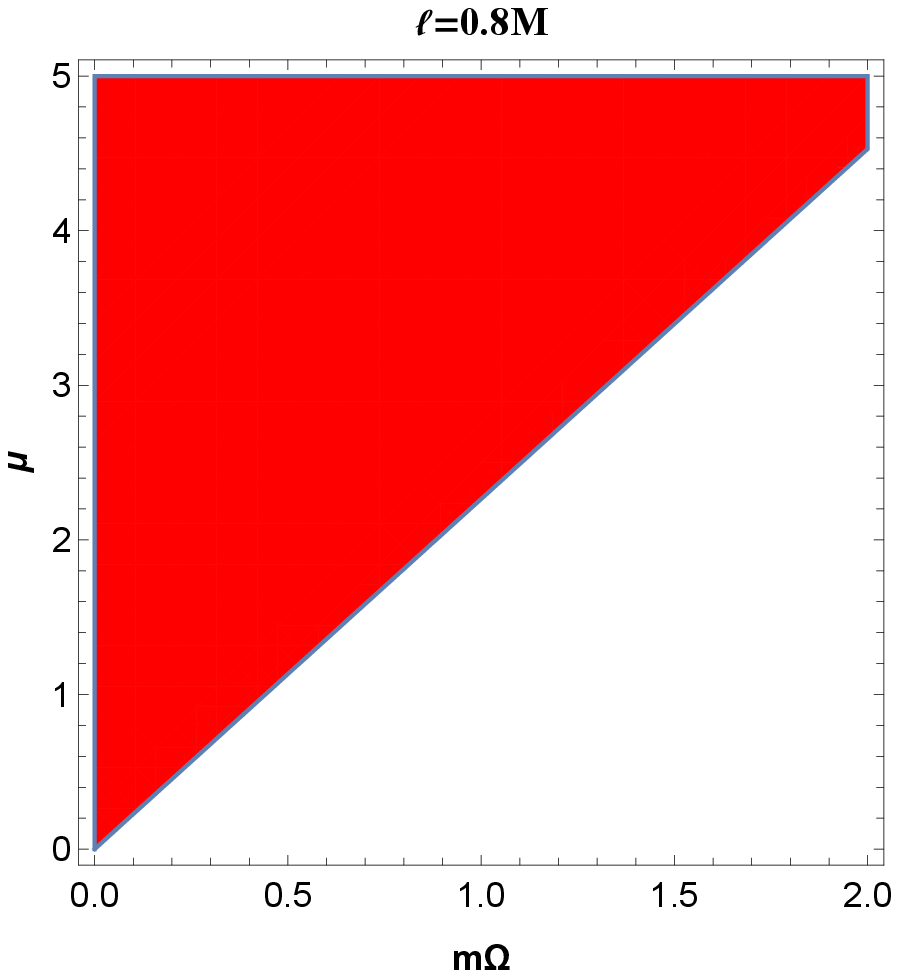}
\end{subfigure}%
\caption{Parameter space($m\Omega$-$\mu$) for massive scalar field
where the colored area represents the region with stable dynamics and
the non-colored area represents the region with unstable dynamics. Here $a=0.52M \quad \text{and}\quad b=0.01M^2$}
\label{fig:test}
\end{figure}
\begin{figure}[H]
\centering
\begin{subfigure}{.34\textwidth}
\includegraphics[width=.9\linewidth]{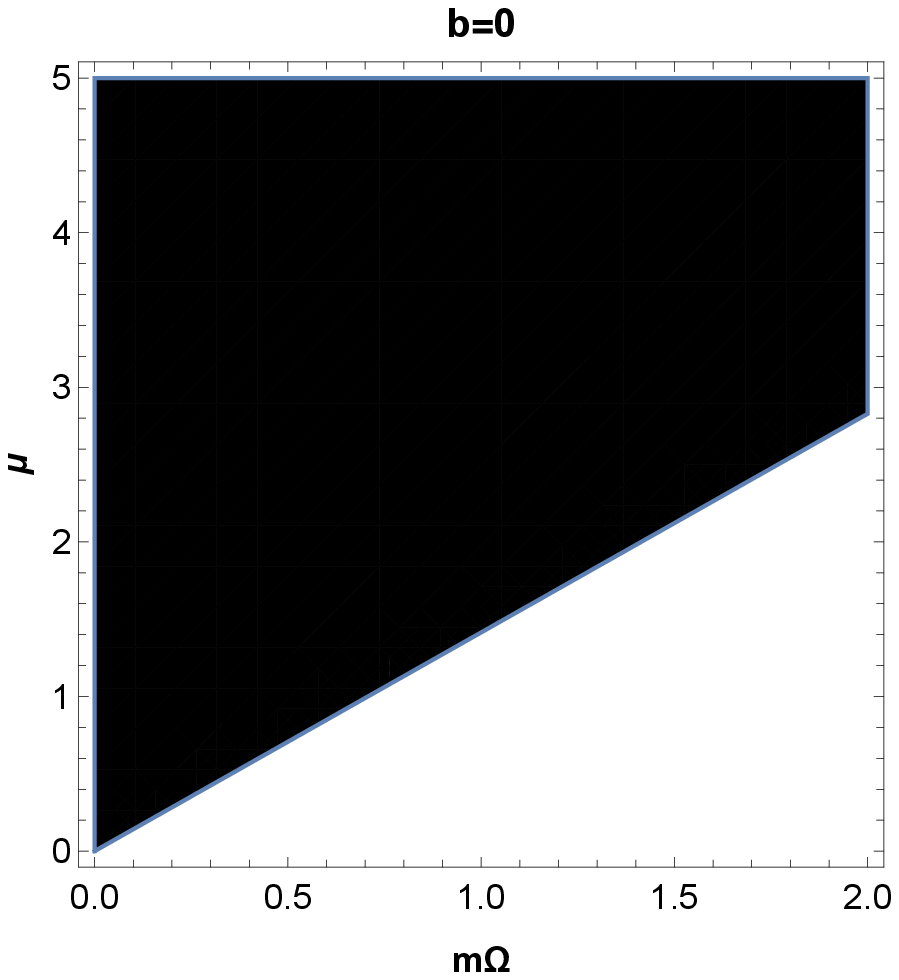}
\end{subfigure}%
\begin{subfigure}{.34\textwidth}
\includegraphics[width=.9\linewidth]{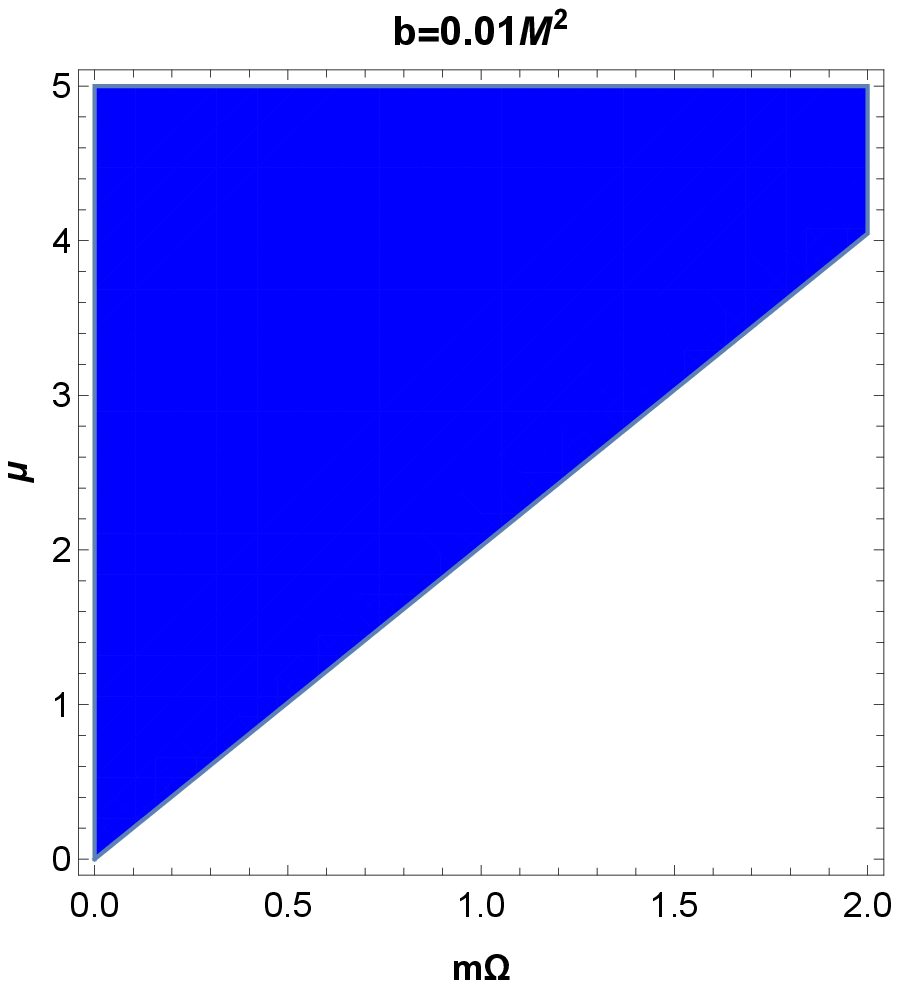}
\end{subfigure}%
\begin{subfigure}{.34\textwidth}
\includegraphics[width=.9\linewidth]{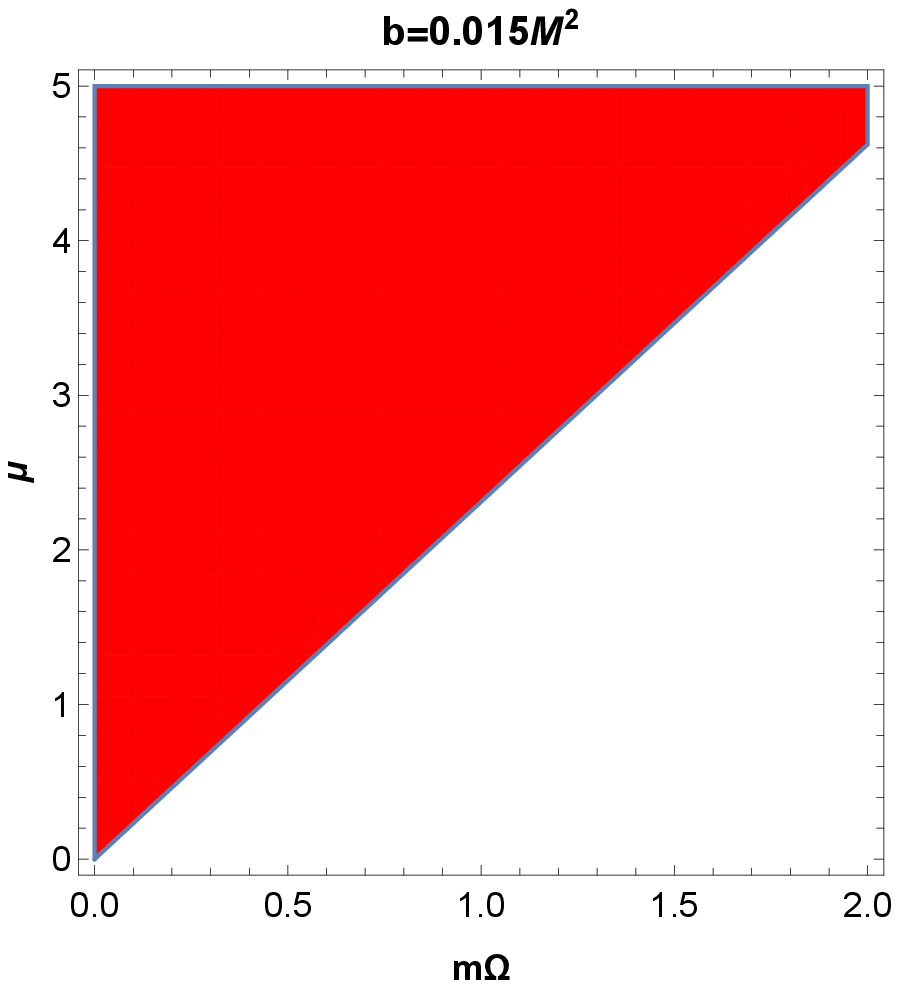}
\end{subfigure}%
\caption{Parameter space($m\Omega$-$\mu$) for massive scalar field
where the colored area represents the region with stable dynamics and
the non-colored area represents the region with unstable dynamics.
Here $a=0.52M\quad \text{and}\quad \ell=0.2M$}
\label{fig:test}
\end{figure}
It can be identified with the Regge-Wheel equation. If we now
discard the terms $\mathcal{O}\left(1 / r^{2}\right)$, the
effective potential $V(r)$ acquires the following asymptotic form
\begin{equation}
V(r)=\mu^{2}-\frac{4 M \omega^{2}}{r}+ \frac{2 M \mu^{2}}{r}.
\label{VAS}
\end{equation}
The potential (\ref{VAS}) will be able to support the trapping of
superradiant waves conspicuously if its asymptotic derivative
becomes positive: $V^{\prime} \rightarrow 0^{+}$ for large $r$,
i.e. when $r \rightarrow \infty$ \cite{HOD}. This together with
the condition of occurrence of superradiance amplification of
scattered waves $\omega < m\Omega_{eh}$, we can mark a specified
regime where $\frac{\mu}{\sqrt{2}} < \omega^{2}<m\Omega_{eh}$.
In this regime, the massive scalar field in the modified
Simpson-Visser background may experience superradiant instability.
This indeed corresponds to the black hole bomb phenomena. However,
the dynamics of the massive scalar field in the extended
Simpson-Visser background will remain stable when
$\mu\geq\sqrt{2}m\Omega_{eh}$
\section{Photon orbit and black hole shadow}
This section is devoted to the study of the black hole shadow
associated with this modified model where the non-commutative
character of spacetime has been amended in the standard
Simpson-Visser background. From the illuminating
studies related to the black hole shadow from which we have got the road map
for the investigation concerning the shadow \cite{SDO1, SDO2, SDO3, SDO4}, we
bring into our consideration two conserved parameters
$\xi$ and $\eta$ as usual to study the shadow. These two are
defined by
\begin{eqnarray}
\xi=\frac{L_{z}}{E} \quad \textrm{and} \quad \eta=\frac{\mathcal{Q}}{E^{2}},
\end{eqnarray}
where $E, L_{z}$, and $\mathcal{Q}$ refers to the energy, the axial component
of the angular momentum, and the Carter constant respectively. In terms of $\xi$
we then express the null geodesics in the Simpson-Visser rotating black hole spacetime:
\begin{eqnarray}\nonumber
\Sigma \frac{d r}{d \lambda}=\pm \sqrt{R}, \quad \Sigma^{2} \frac{d
\theta}{d \lambda}=\pm \sqrt{\Theta}, \\\nonumber
\Delta \Sigma \frac{d t}{d \lambda}=A-2\operatorname{M\sqrt{r^2+\ell^2}a\xi},\\
\Delta \Sigma \frac{d \phi}{d \lambda}=2M\sqrt{r^2+\ell^2} a+\frac{\xi}{\sin ^{2}
\theta}\left(\rho^{2}-2 M\sqrt{r^2+\ell^2}\right),
\end{eqnarray}
where $\lambda$ is the affine parameter and $R(r)$ has the expression
\begin{eqnarray}
R(r)=\left[r^{2}+\ell^2+a^{2}-a
\xi\right]^{2}-\Delta\left[\eta+(\xi-a)^{2}\right],\quad
\Theta(\theta)=\eta+a^{2} \cos ^{2} \theta-\xi^{2} \cot
^{2} \theta.
\end{eqnarray}
Now, the radial equation of motion can be written down in the
familiar form:
\begin{eqnarray}
\left(\Sigma \frac{d r}{d \lambda}\right)^{2}+V_{e f f}=0.
\label{VEF}
\end{eqnarray}
The effective potential $V_{e f f}$ in Eq. (\ref{VEF}) has the
expression
\begin{eqnarray}
V_{e f f}=-\left[r^{2}+\ell^2+a^{2}-a
\xi\right]^{2}+\Delta\left[\eta+(\xi-a)^{2}\right].
\end{eqnarray}
The unstable spherical orbit on the equatorial plane,
$\theta=\frac{\pi}{2}$ is described by the equation
\begin{eqnarray}
R(r)=0,\quad \frac{d R}{d r}=0, \label{CONDITION1}
\end{eqnarray}
with conditions $ \frac{d^{2} R}{d r^{2}} < 0, \quad
\textrm{and} \quad \eta=0$.

For more generic orbits $\theta \neq \pi / 2$, and $\eta \neq 0$,
the solution of Eqn. (\ref{CONDITION1})$r=r_{s}$ gives the $r-$
constant orbit, which is also called spherical orbit, and the
conserved parameters of the spherical orbits read
\begin{eqnarray}
\xi_{s}&=&\frac{\left(a^{2}+\ell^2+r^{2}\right)\Delta^{'}(r)
-4\sqrt{r^2+\ell^2}\Delta(r)}{a\Delta^{'}(r)},\\\nonumber
\eta_{s}&=&\frac{\left(8r\Delta(r)\left(2a^{2}r
+(r^2+\ell^2)\Delta^{'}(r)\right)-(r^{2}+\ell^2)^2\Delta^{'}(r)^{2}
-16r^2\Delta(r)^{2}\right)}{a^{2}\Delta^{'}(r)^{2}},
\end{eqnarray}
where upper prime stands for differentiation with respect to
radial coordinate. The above expressions reduce to the
corresponding items for Kerr black hole when both $\ell$ and $b$
approach a vanishing value. It is helpful at this stage to launch
two celestial coordinates for studying the shadow in a preferred
manner. The celestial coordinates, which are used here to
characterize the shadow which
would be expected to be viewed by an observer in the sky, could be given by
\begin{eqnarray}\nonumber
\alpha(\xi, \eta ; \theta)&=&\lim _{r \rightarrow \infty} \frac{-r
p^{(\varphi)}}{p^{(t)}} = -\xi_{s} \csc \theta,\\\nonumber
\beta(\xi, \eta ; \theta)&=&\lim _{r \rightarrow \infty} \frac{r
p^{(\theta)}}{p^{(t)}}
=\sqrt{\left(\eta_{s}+a^{2} \cos ^{2} \theta-\xi_{s}^{2} \cot ^{2} \theta\right)},\\
\end{eqnarray}
where $\left(p^{(t)}, p^{(r)}, p^{(\theta)}, p^{(\phi)}\right)$
are the tetrad components of the momentum of the photon with
respect to a locally non-rotating reference frame \cite{BARDEEN}.
With these inputs, the stage is now set for sketching the shadows
of this black hole for various cases which are portrayed in the
following Figs. (11, 12).
\begin{figure}[H]
\centering
\begin{subfigure}{.5\textwidth}
\centering
\includegraphics[width=.7\linewidth]{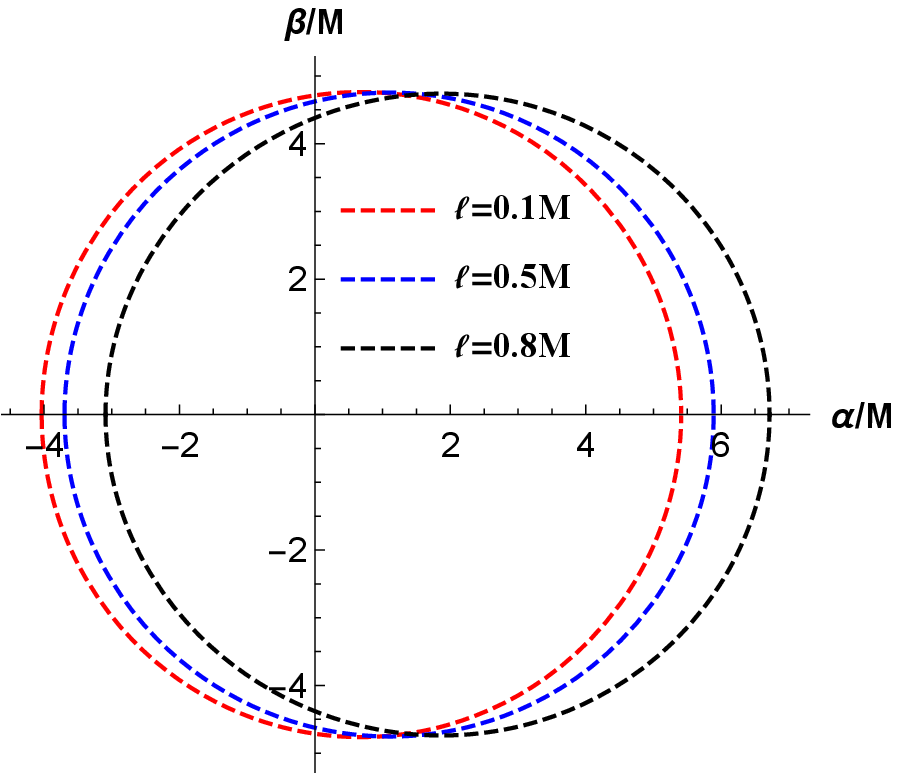}
\end{subfigure}%
\begin{subfigure}{.5\textwidth}
\centering
\includegraphics[width=.7\linewidth]{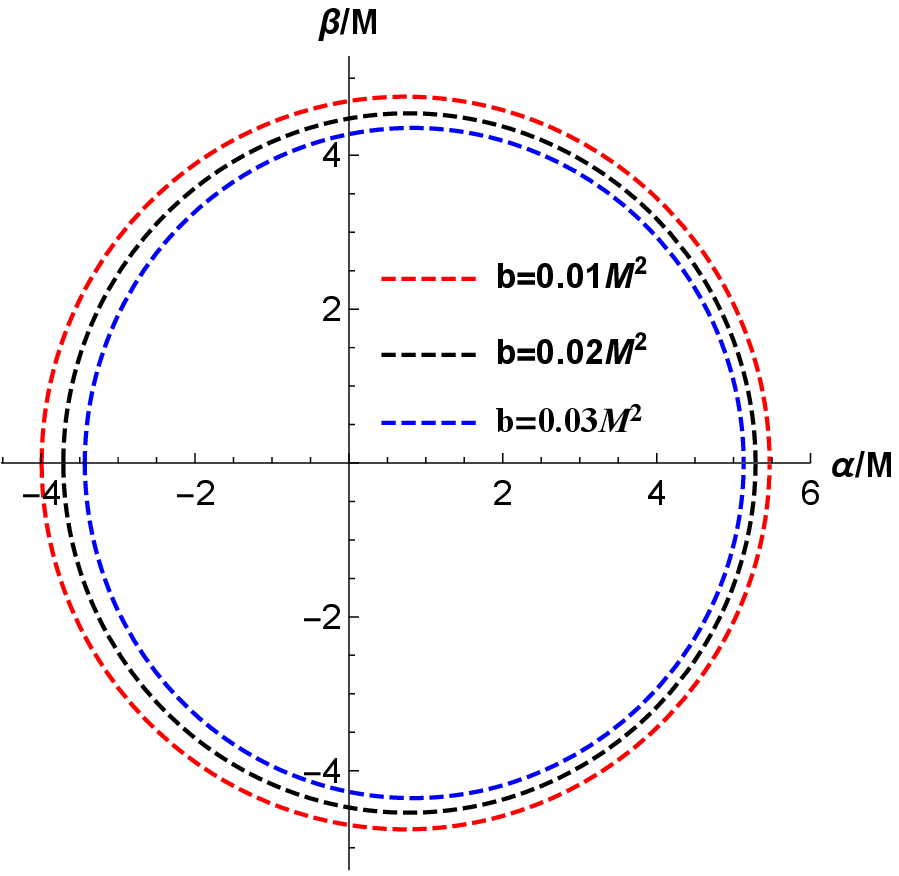}
\end{subfigure}%
\caption{The left panel gives shapes of the shadow for various
values of $\ell$ with $a=0.3M$, $b=0.01M^{2}$, and $\theta=\pi/2$.
The right panel gives shapes of the shadow for various values of
$b$ with $a=0.3M$ , $\ell=0.2M$, and $\theta=\pi/2$.} \label{fig:test}
\end{figure}

\begin{figure}[H]
\centering
\begin{subfigure}{.5\textwidth}
\centering
\includegraphics[width=.7\linewidth]{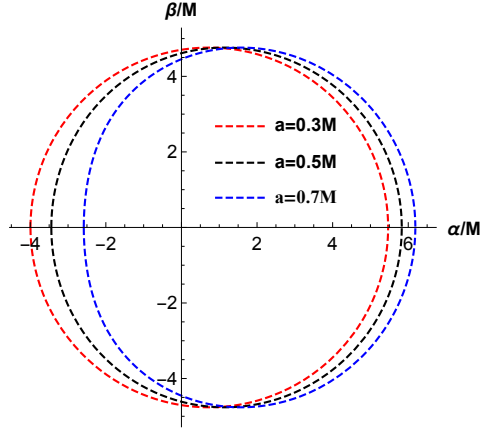}
\end{subfigure}
\caption{ The shapes of the shadow for various values of $a$ with
$b=0.01M^{2}$, $\ell=0.2M$, and $\theta=\pi/2$. } \label{fig:test}
\end{figure}
A careful look at the plots reveals that the geometrical size
of the shadow decreases with an increase in the value of $b$.
Furthermore, one can also observe that the left side of the shadow
gets shifted towards the right if $\ell$ or $a$ increases.

Let us now introduce the parameters defined by Hioki and Maeda \cite{KH} to analyze the deviation
from the circularity
$\left(\delta_{s}\right)$ and the size $\left(R_{s}\right)$ of
the shadow cast by the black hole.
\begin{figure}[H]
\centering
\begin{subfigure}{.5\textwidth}
\centering
\includegraphics[width=.7\linewidth]{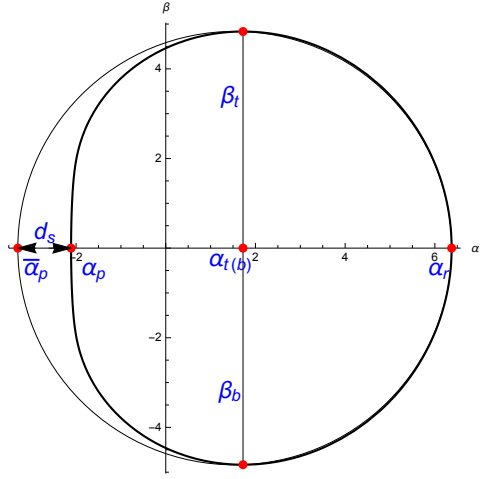}
\end{subfigure}
\caption{The black hole shadow and reference circle. $ds$ is the
distance between the left points of the shadow and the reference
circle.} \label{fig:test}
\end{figure}
To compute these parameters, we consider five reference points
$\left(\alpha_{t}, \beta_{t}\right),\left(\alpha_{b},
\beta_{b}\right),\left(\alpha_{r}, 0\right)$ $\left(\alpha_{p},
0\right)$ and $\left(\bar{\alpha}_{p}, 0\right)$. Out of these
five points, the first four are describing the top, bottom,
rightmost, and leftmost points of the shadow respectively, and the
last one refers to the leftmost point of the reference circle in
Fig.~(13). Therefore, from the geometry we can write

$$
R_{s}=\frac{\left(\alpha_{t}-\alpha_{r}\right)^{2}+\beta_{t}^{2}}{2\left|\alpha_{t}-\alpha_{r}\right|}
$$
and
$$
\delta_{s}=\frac{\left|\bar{\alpha}_{p}-\alpha_{p}\right|}{R_{s}}.
$$
In the following Figs.~(14, 15, 16, 17) we plot $R_{s}$ and
$\delta_{s}$ for various cases to examine how $R_{s}$ and
$\delta_{s}$ vary with parameters involved in this modified
Simpson-Visser model of gravity.
\begin{figure}[H]
\centering
\begin{subfigure}{.5\textwidth}
\centering
\includegraphics[width=.7\linewidth]{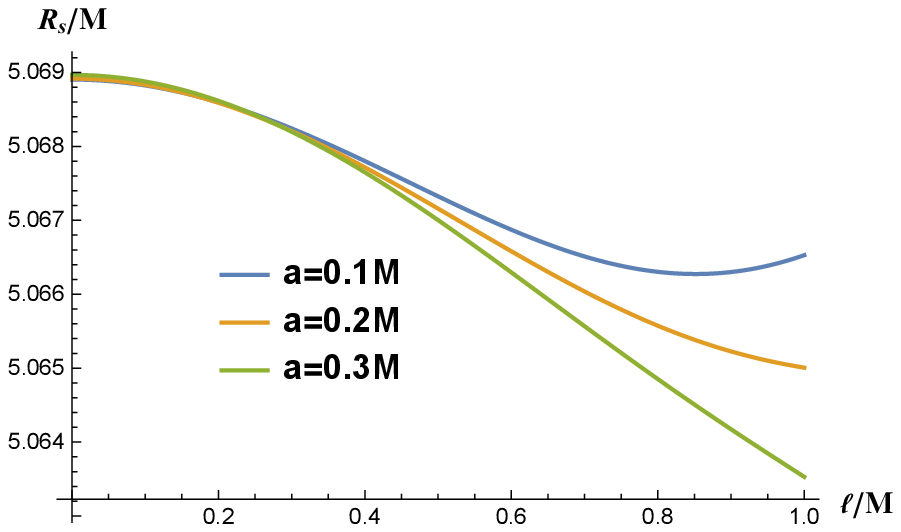}
\end{subfigure}%
\begin{subfigure}{.5\textwidth}
\centering
\includegraphics[width=.7\linewidth]{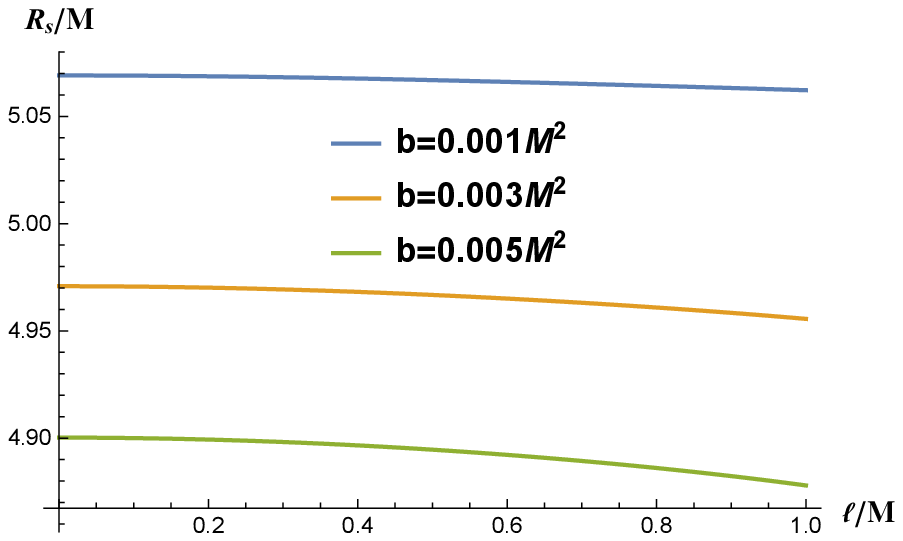}
\end{subfigure}
\caption{The left panel shows variation of $R_{s}$ with respect to $\ell$ for various
values of $a$ with $b=.001M^2$ and $\theta=\pi/2$. The right
panel shows the variation of $R_{s}$ with respect to $\ell$ for various values of $b$
with $a=0.4M$ and $\theta=\pi/2$. } \label{fig:test}
\end{figure}

\begin{figure}[H]
\centering
\begin{subfigure}{.5\textwidth}
\centering
\includegraphics[width=.7\linewidth]{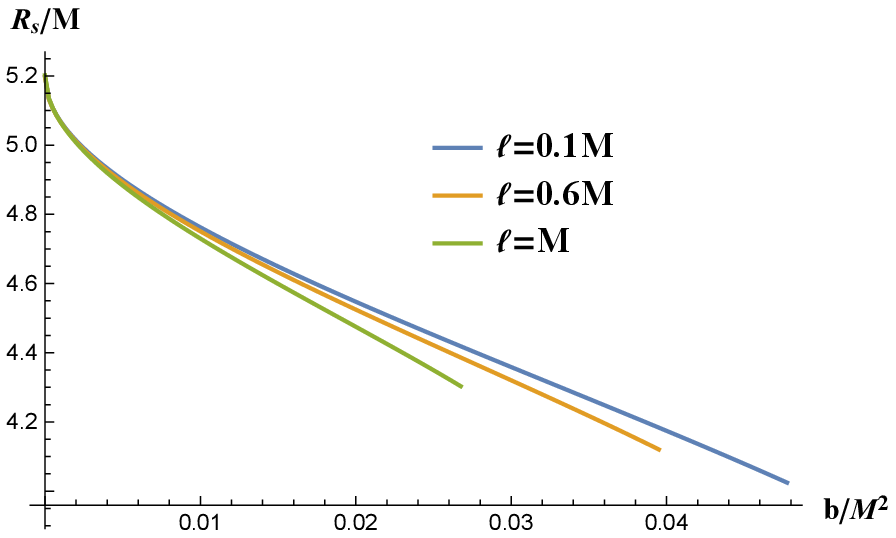}
\end{subfigure}
\caption{The left panel shows variation of $R_{s}$  with respect to $b$ for various
values of $\ell$ with $a=0.1M$ and $\theta=\pi/2$. } \label{fig:test}
\end{figure}

\begin{figure}[H]
\centering
\begin{subfigure}{.5\textwidth}
\centering
\includegraphics[width=.7\linewidth]{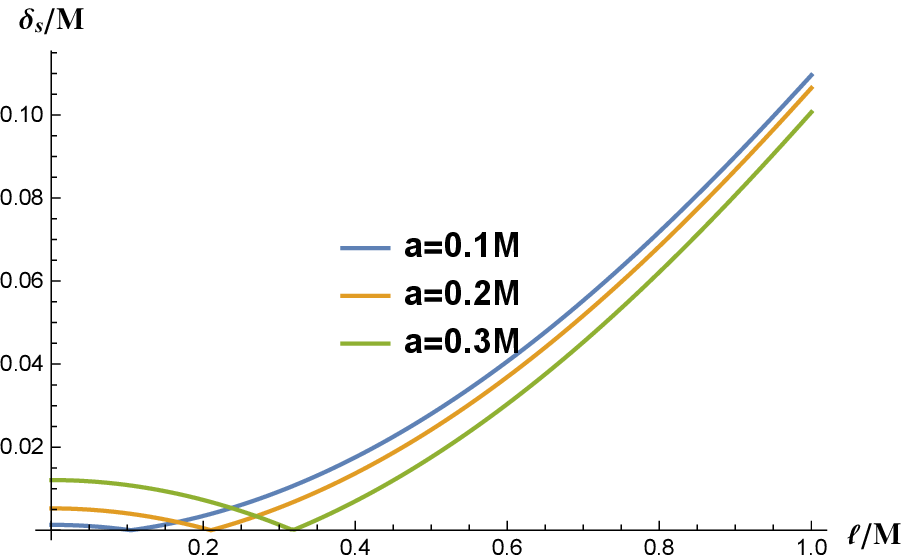}
\end{subfigure}%
\begin{subfigure}{.5\textwidth}
\centering
\includegraphics[width=.7\linewidth]{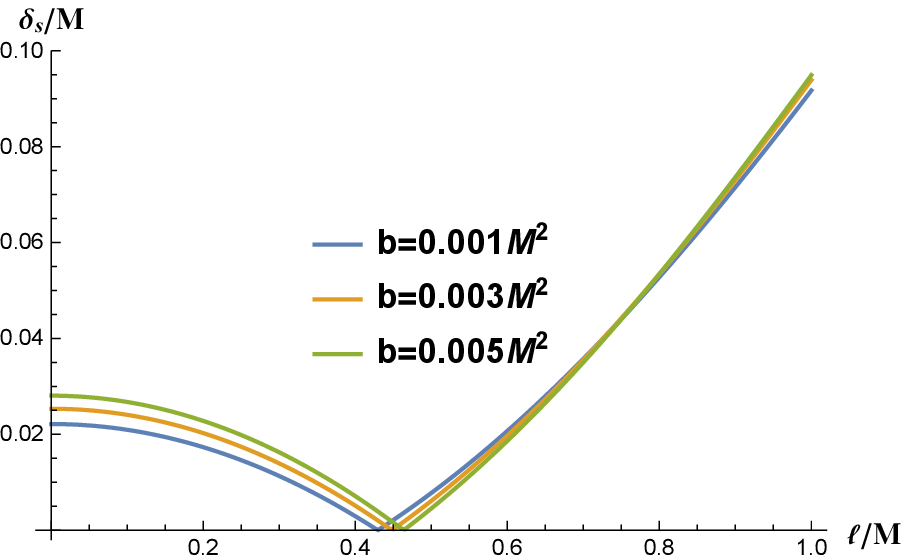}
\end{subfigure}
\caption{The left panel shows variation of $\delta_{s}$ for various
values of $a$ with $b=.001M^2$, $\theta=\pi/2$, and the right
panel shows the variation of $\delta_{s}$ for various values of $b$
with $a=0.4M$ and $\theta=\pi/2$.} \label{fig:test}
\end{figure}

\begin{figure}[H]
\centering
\begin{subfigure}{.5\textwidth}
\centering
\includegraphics[width=.7\linewidth]{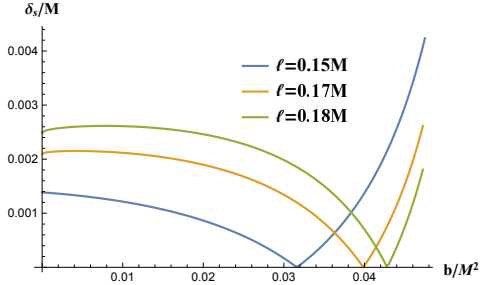}
\end{subfigure}
\caption{The left panel shows variation of $R_{s}$ for various
values of $\ell$ with $a=0.1M$ and $\theta=\pi/2$.} \label{fig:test}
\end{figure}
From the above plots in Figs. (14, 15) we observe that $R_{s }$
decreases with an increase in $b$ for fixed values of $a$ and
$\ell$. It is also found that $R_{s}$ decreases for an
increase in the value of $a$, keeping $b$ and $\ell$ fixed or an increase in $\ell$ for fixed values of $a$ and $b$. Furthermore,
it is found in Figs. (16, 17) that $\delta_{s}$ initially
decreases and then increases with an increase in the value of
either $\ell$ or $b$. For some specific combination of values of
the parameters $a$, $b$, and $\ell$ we may get $\delta_{s}=0$. One
such combination is $b=0.001$, $\ell=0.210082$, and $a=0.2M$. It
signifies that for such a combination of parameters the shadow
becomes a perfect circle.
\section{Constraining from the observed data for $\mathrm{M}87^{*}$}
In this section, an attempt is made towards constraining the
parameters associated with this modified theory. We compare the
geometry of the shadow obtained from the numerical
computation for the
Simpson-Visser black hole with the non-commutative setting with
the geometry of the observed shadow for the
$\mathrm{M}87^{*}$ black hole. We envisage the experimentally
obtained astronomical data corresponding to deviation from
circularity $\Delta \leq 0.10$ and angular diameter
$\theta_{d}=42\pm 3 \mu as$. The boundary of the shadow is described by the polar coordinates
$(R(\phi),\phi)$ with the origin lays at the center of the shadow
$(\alpha_{C}, \beta_{C})$ where
$\alpha_{C}=\frac{|\alpha_{max}+\alpha_{min}|}{2}$, and
$\beta_{C}=0$. If a point $(\alpha, \beta)$ is considered over the
boundary of the image which subtends an angle $\phi$ on the
$\alpha$ axis at the geometric center ($\left(\alpha_{C},
0\right)$) and the distance between the points $(\alpha, \beta)$
$\left(\alpha_{C}, 0\right)$ is marked by $R(\phi)$, then the
average radius $R_{\text{avg}}$ of the image is given by
\cite{CBK}
\begin{equation}
R_{\text {avg}}^{2} \equiv \frac{1}{2 \pi} \int_{0}^{2 \pi} d \phi
R^{2}(\phi), \label{RAV}
\end{equation}
where $R(\phi) \equiv \sqrt{\left(\alpha(\phi)-\alpha_{C}\right)^{2}+\beta(\phi)^{2}}$,
and $\phi = tan^{-1}\frac{\beta(\phi)}{\alpha(\phi)-\alpha_{C}}$.

With this geometrical information, we have the definition of
deviation from circularity $\Delta C$ \cite{TJDP}:
\begin{equation}
\Delta C \equiv 2\sqrt{\frac{1}{2 \pi} \int_{0}^{2 \pi} d
\phi\left(R(\phi)-R_{\text {avg }}\right)^{2}}, \label{DELTAC}
\end{equation}
and the angular diameter of the shadow which is defined by
\begin{equation}
\theta_{d}=\frac{2}{d}\sqrt{\frac{A}{\pi}}, \label{THETA}
\end{equation}
where $A=2\int_{r_{-}}^{r_{+}} \beta d\alpha $ is the area of the
shadow. For $M87^{*}$ black hole, the distance of $M87^{*}$
from the Earth is $d=16.8 Mpc$. Equations (\ref{RAV}, \ref{DELTAC},\ref{THETA} ) along with the available data
for $M87^{*}$ enable us to accomplish a comparison of the
theoretical predictions for modified Simpson-Visser black-hole
shadows with the experimental findings provided by the EHT
collaboration. In Figs. (18, 19), the deviation from circularity,
$\Delta C$ is shown for the extended Simpson-Visser black holes
for the two specified inclination angles $\theta=90^{o}$ and
$\theta=17^{o}$ respectively.
\begin{figure}[H]
\centering
\begin{subfigure}{.25\textwidth}
\centering
\includegraphics[scale=.65]{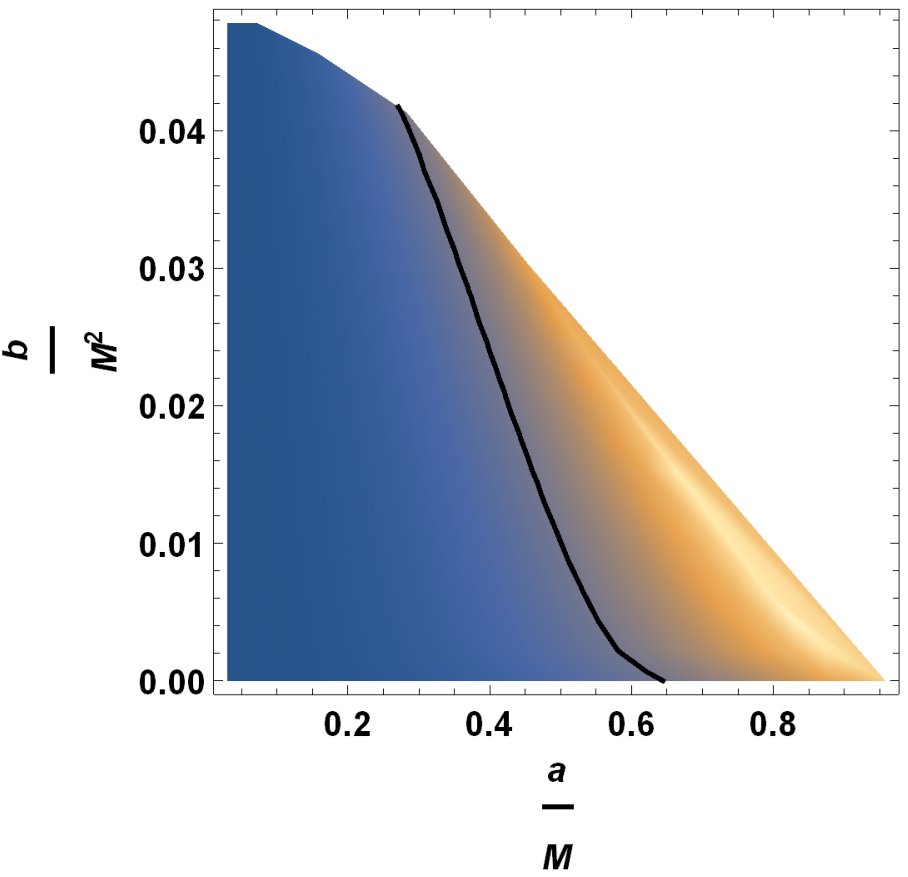}
\end{subfigure}%
\begin{subfigure}{.28\textwidth}
\centering
\raisebox{.2\height}{\includegraphics[scale=.55]{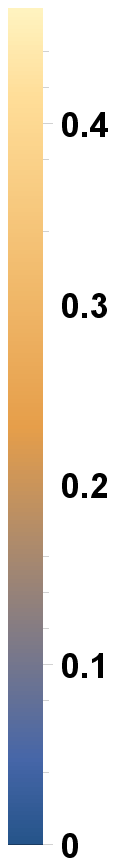}}\hspace{1.5em}%
\end{subfigure}%
\begin{subfigure}{.25\textwidth}
\centering
\includegraphics[scale=.65]{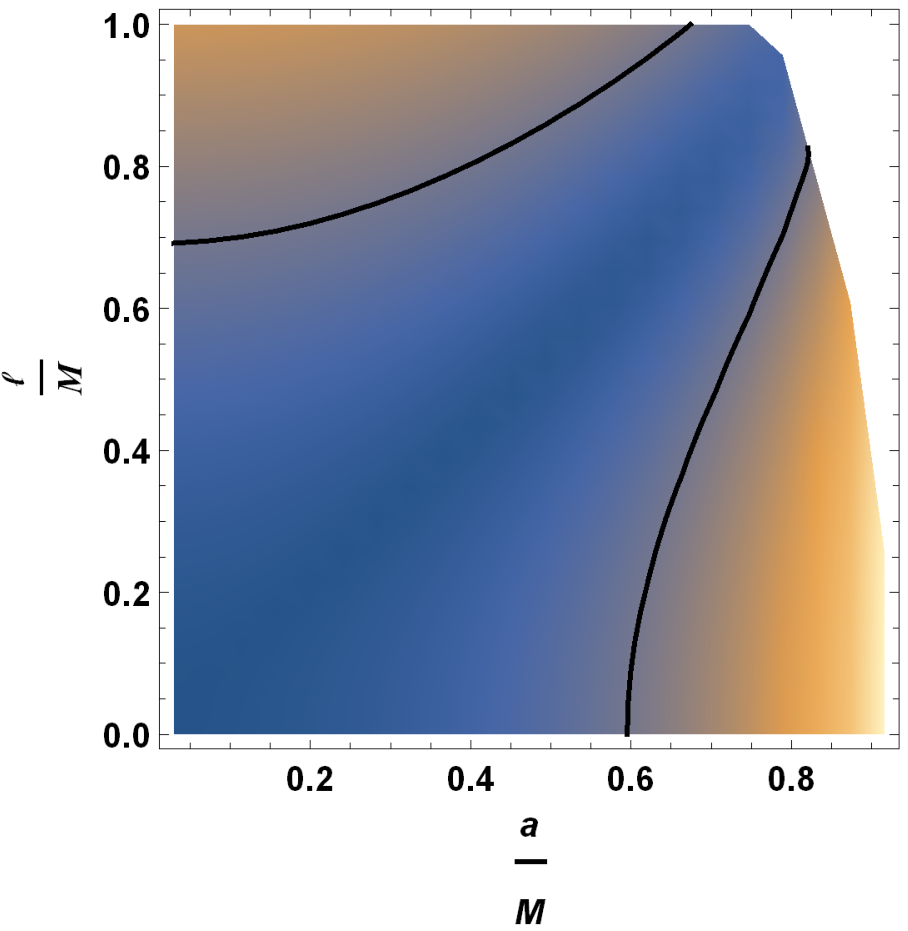}
\end{subfigure}%
\begin{subfigure}{.28\textwidth}
\centering
\raisebox{.19\height}{\includegraphics[scale=.6]{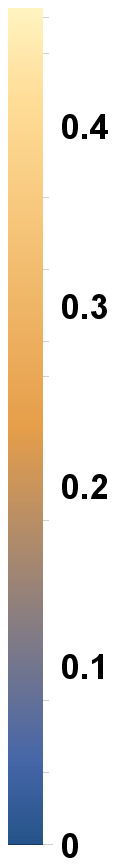}}
\end{subfigure}
\caption{The left panel is for $\ell=0.1M$ and the right panel is for
$b=0.001M^{2}$ where the inclination angle is $90^{o}$. The black
solid lines correspond to $\Delta C=0.1$. }
\end{figure}
\smallskip
\begin{figure}[H]
\centering
\begin{subfigure}{.25\textwidth}
\centering
\includegraphics[scale=.65]{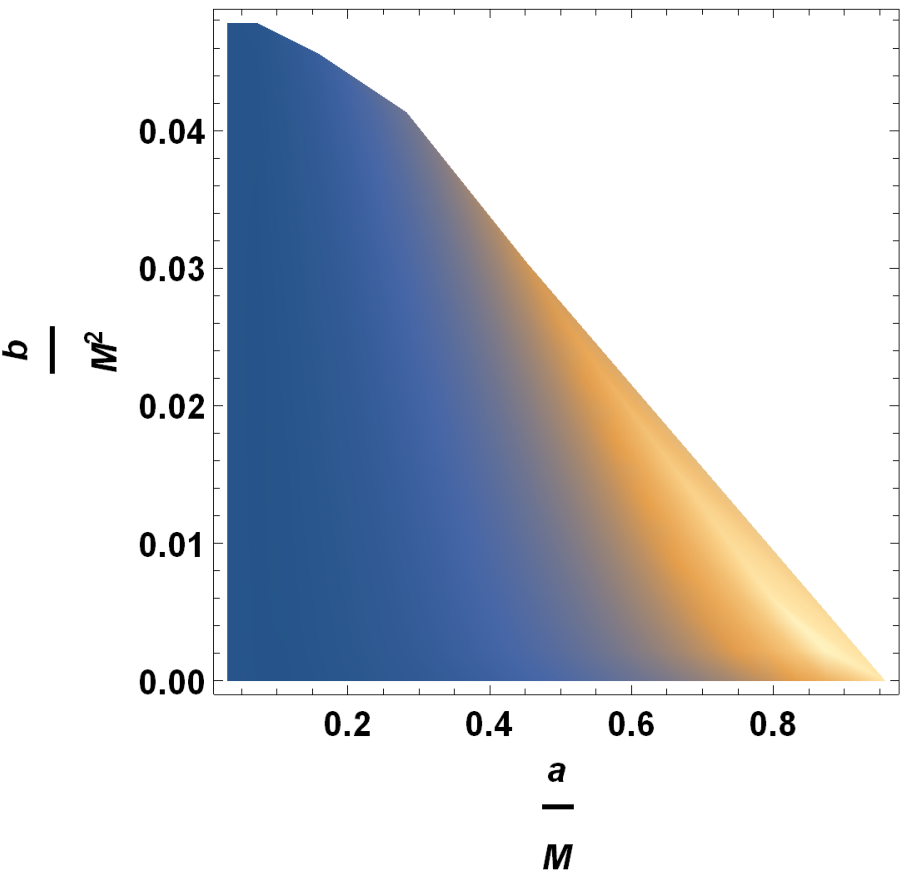}\hspace{1.5em}%
\end{subfigure}%
\begin{subfigure}{.28\textwidth}
\centering
\raisebox{.2\height}{\includegraphics[scale=.55]{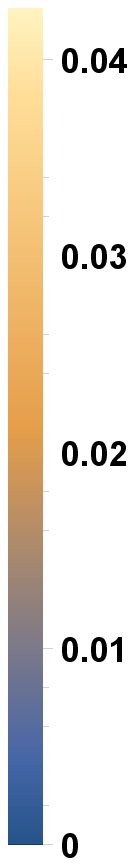}}\hspace{1.5em}%
\end{subfigure}%
\begin{subfigure}{.25\textwidth}
\centering
\includegraphics[scale=.65]{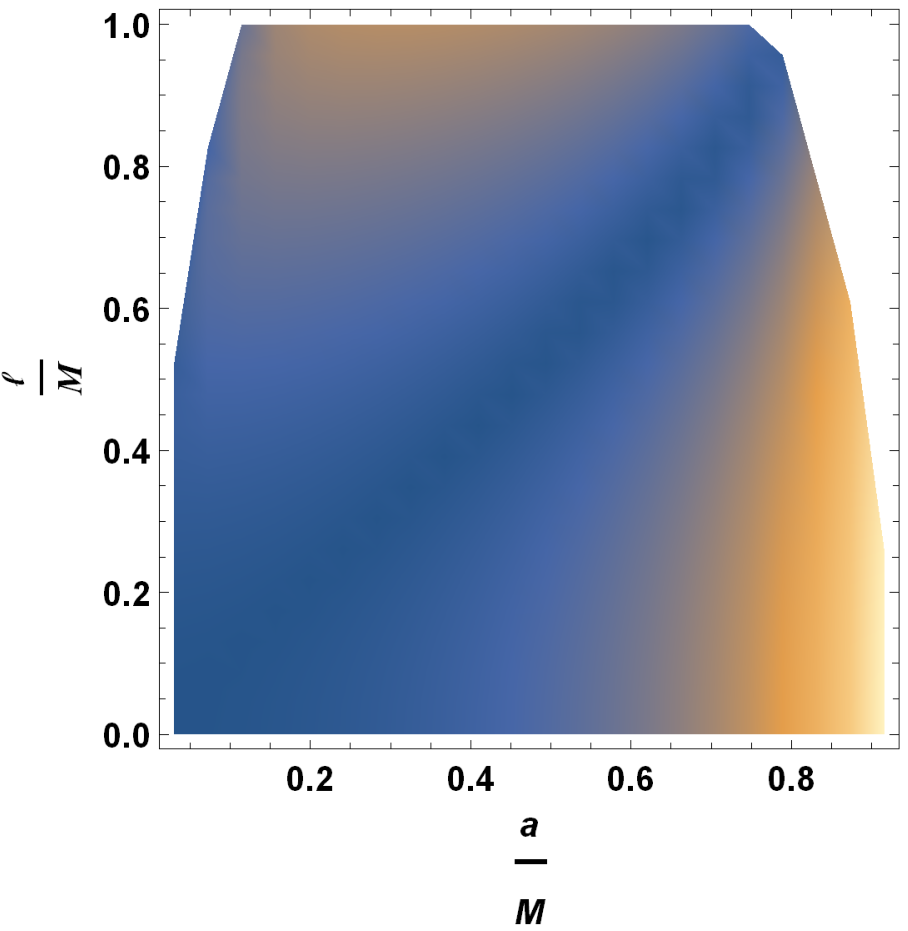}\hspace{1.5em}%
\end{subfigure}%
\begin{subfigure}{.28\textwidth}
\centering
\raisebox{.19\height}{\includegraphics[scale=.6]{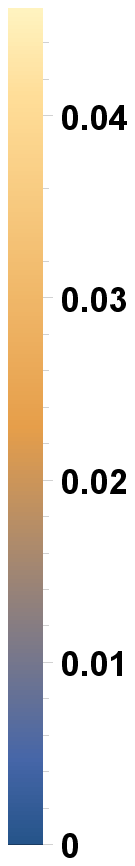}}
\end{subfigure}
\caption{The left panel is for $\ell=0.1M$ and the right one is for
$b=0.001M^{2}$ where the inclination angle is $17^{o}$.}
\end{figure}
In figs. (20, 21) the angular diameter $\theta_{d}$ is shown
for non-commutative Simoson-Visser black holes for inclination
angles $\theta=90^{o}$ and $\theta=17^{o}$ respectively.

\begin{figure}[H]
\centering
\begin{subfigure}{.25\textwidth}
\centering
\includegraphics[scale=.65]{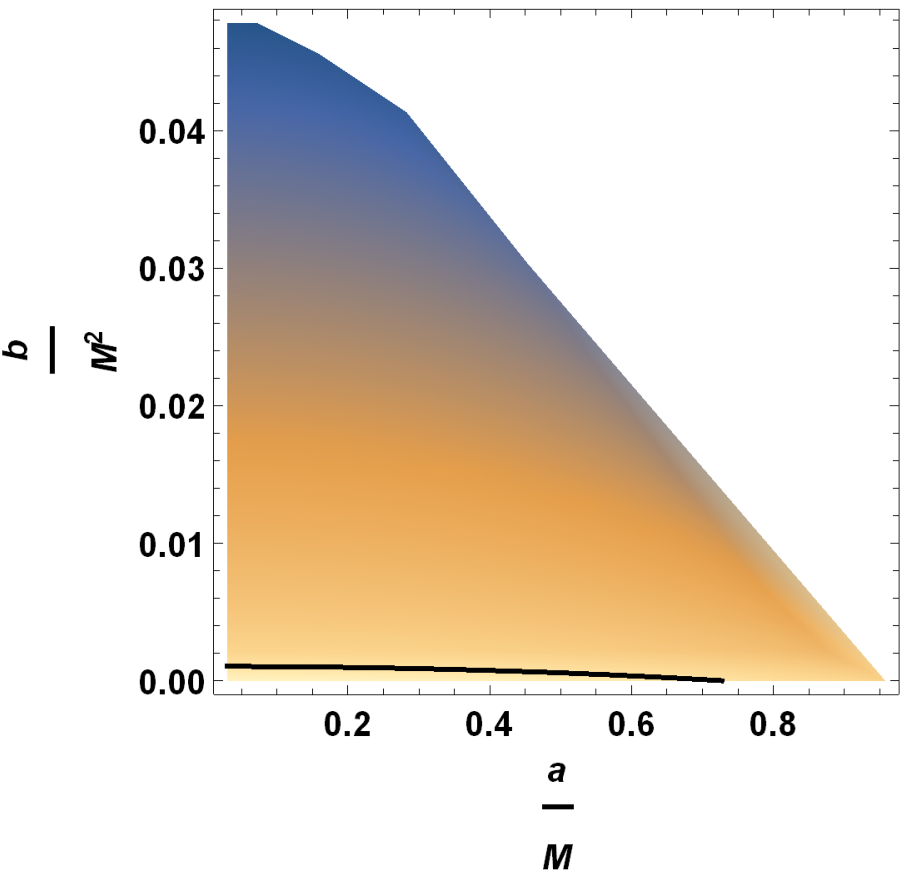}\hspace{1.5em}%
\end{subfigure}%
\begin{subfigure}{.28\textwidth}
\centering
\raisebox{.2\height}{\includegraphics[scale=.55]{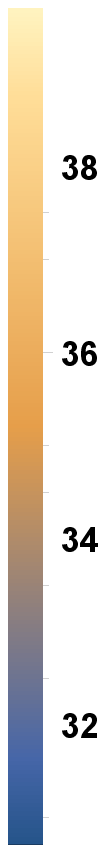}}\hspace{1.5em}%
\end{subfigure}%
\begin{subfigure}{.25\textwidth}
\centering
\includegraphics[scale=.65]{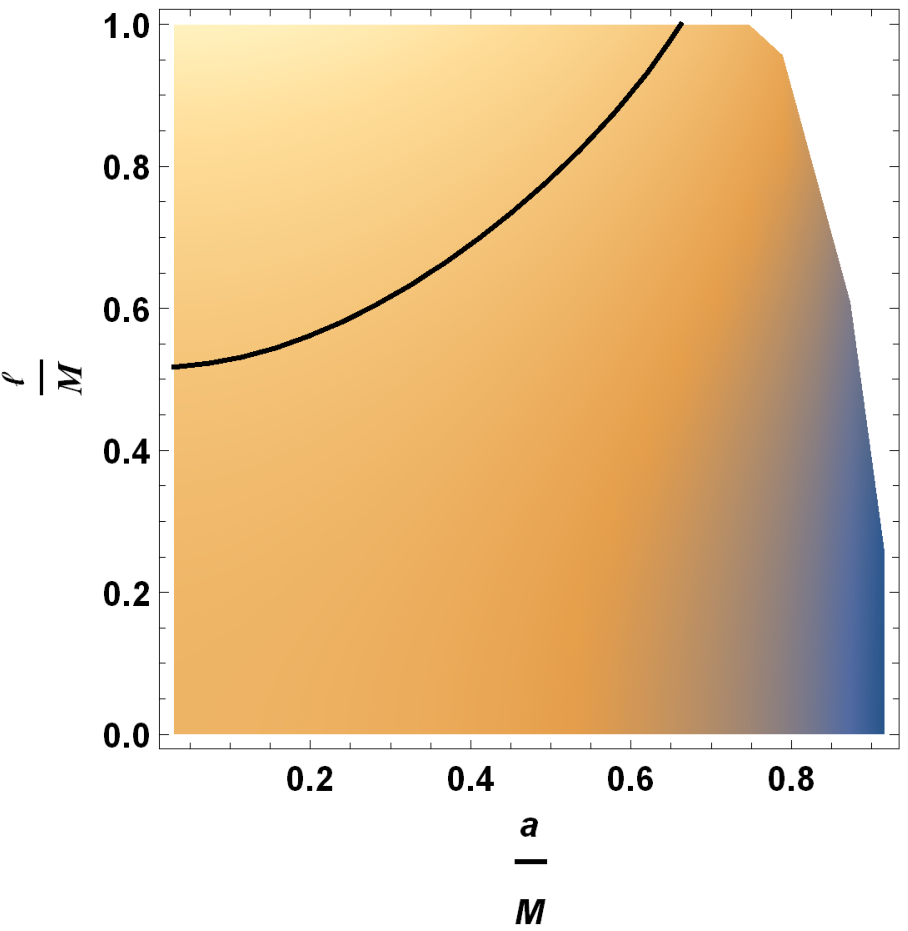}\hspace{1.5em}%
\end{subfigure}%
\begin{subfigure}{.28\textwidth}
\centering
\raisebox{.19\height}{\includegraphics[scale=.6]{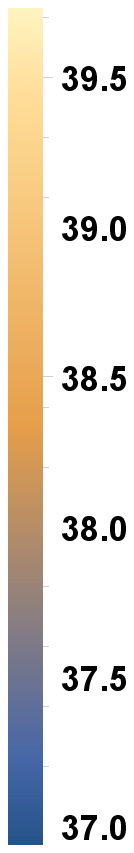}}
\end{subfigure}
\caption{The left panel is for $\ell=0.1M$ and the right panel is for
$b=0.001M^{2}$ where the inclination angle is $90^{o}$. The black
solid lines correspond to $\theta_{d}=39 \mu as$.}
\end{figure}
\smallskip
\begin{figure}[H]
\centering
\begin{subfigure}{.25\textwidth}
\centering
\includegraphics[scale=.65]{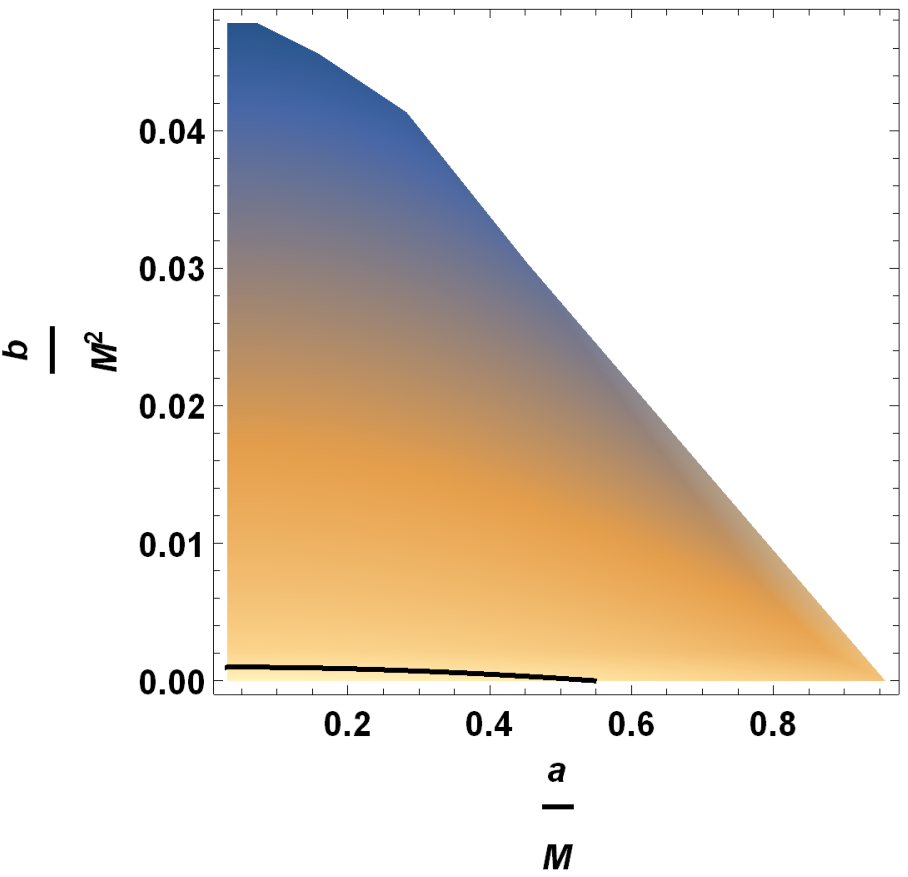}\hspace{1.5em}%
\end{subfigure}%
\begin{subfigure}{.28\textwidth}
\centering
\raisebox{.2\height}{\includegraphics[scale=.55]{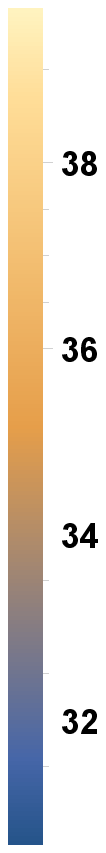}}\hspace{1.5em}%
\end{subfigure}%
\begin{subfigure}{.25\textwidth}
\centering
\includegraphics[scale=.65]{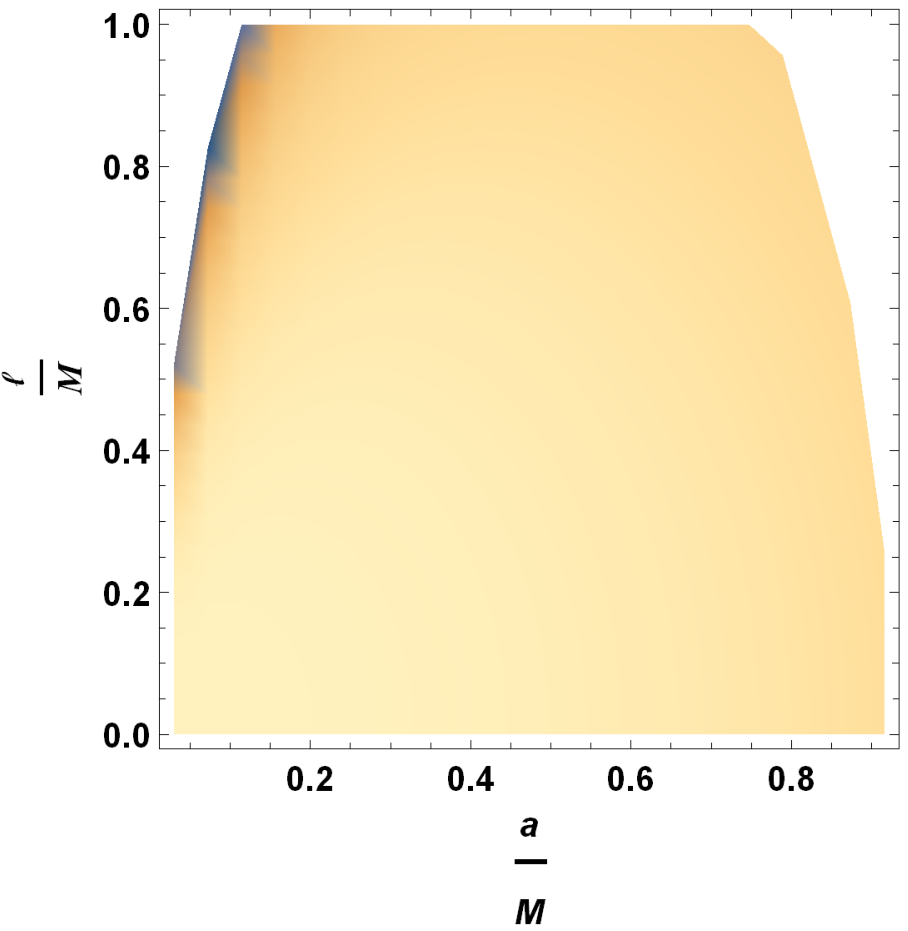}\hspace{1.5em}%
\end{subfigure}%
\begin{subfigure}{.28\textwidth}
\centering
\raisebox{.19\height}{\includegraphics[scale=.6]{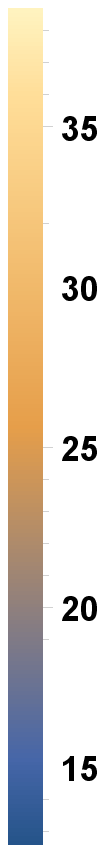}}
\end{subfigure}
\caption{The left panel is for $\ell=0.1M$ and the right panel is for
$b=0.001M^{2}$ where the inclination angle is $17^{o}$. The black
solid lines correspond to $\theta_{d}=39 \mu as$.}
\end{figure}
From Figs. (18, 19), it transpires that the constraint $\Delta
C\leq 0.1$ is satisfied for finite parameter space when the
inclination angle is $90^{o}$, but for $17^{o}$, the constrain is
satisfied for the entire parameter space. Figs.(20, 21) reveal
that for the inclination angles $\theta=90^{o}$ and
$\theta=17^{o}$, the constraint $\theta_{d}=42\pm3\mu$ is satisfied for a finite parameter space within the $1\sigma$ region.
The asymmetry from the exact circular nature in the $M87^{*}$
shadow can also be determined in terms of the axial ratio $D_{X}$
which is the ratio of the major to the minor diameter of the
shadow \cite{KA1}. The axial ratio $D_{X}$ is defined by
\cite{BCS}
\begin{equation}
D_{X}=\frac{\Delta Y}{\Delta X}=\frac{\beta_{t}-\beta_{b}}{\alpha_{r}-\alpha_{p}}.
\end{equation}
We should have $D_{X}$ within the range $1 < D_{X}\lesssim4/3$
according to the EHT observations associated with $M87^{*}$ black
hole \cite{KA1}. In fact, $D_{X}$ is defined to determine $\Delta
C$ in a different manner. The measured axial ratio for $M87^{*}$
is $4:3$ which is in good agreement to $\Delta C \leq 0.1$
\cite{KA1}. In the Figs. below, axial ratio $D_{X}$ is shown for
the non-commutative Simpson-Visser black hole for inclination
angles $\theta=90^{o}$ and $\theta=17^{o}$ respectively.

\begin{figure}[H]
\centering
\begin{subfigure}{.25\textwidth}
\centering
\includegraphics[scale=.65]{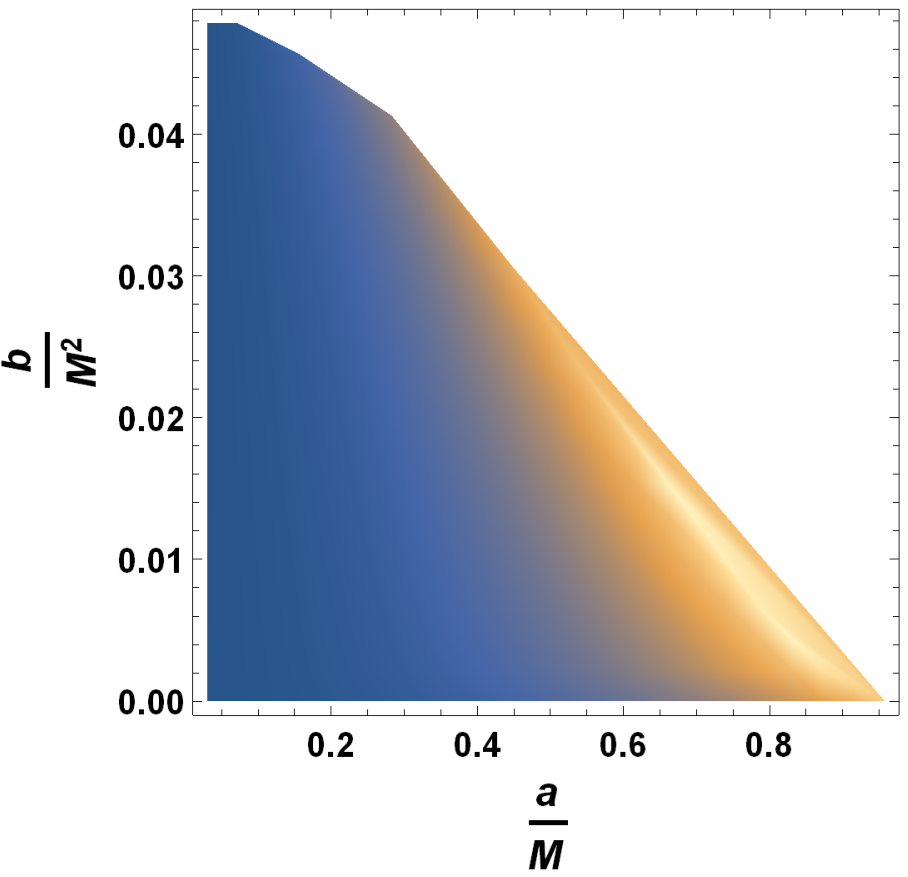}\hspace{1.5em}%
\end{subfigure}%
\begin{subfigure}{.28\textwidth}
\centering
\raisebox{.2\height}{\includegraphics[scale=.55]{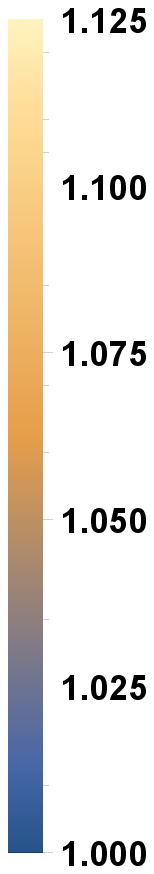}}\hspace{1.5em}%
\end{subfigure}%
\begin{subfigure}{.25\textwidth}
\centering
\includegraphics[scale=.65]{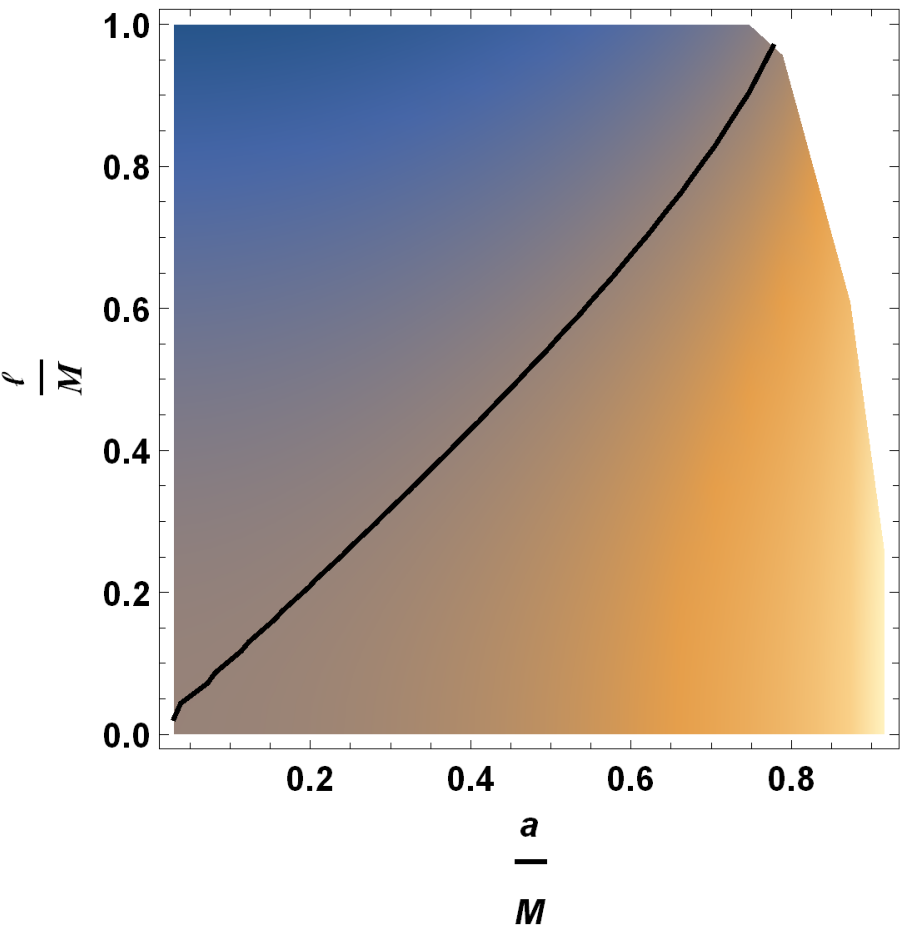}\hspace{1.5em}%
\end{subfigure}%
\begin{subfigure}{.28\textwidth}
\centering
\raisebox{.19\height}{\includegraphics[scale=.6]{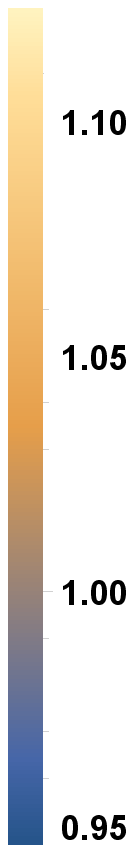}}
\end{subfigure}
\caption{The left panel is for $\ell=0.1M$ and the right panel is for
$b=0.001M^{2}$ where the inclination angle is $90^{o}$.}
\end{figure}
\smallskip
\begin{figure}[H]
\centering
\begin{subfigure}{.25\textwidth}
\centering
\includegraphics[scale=.65]{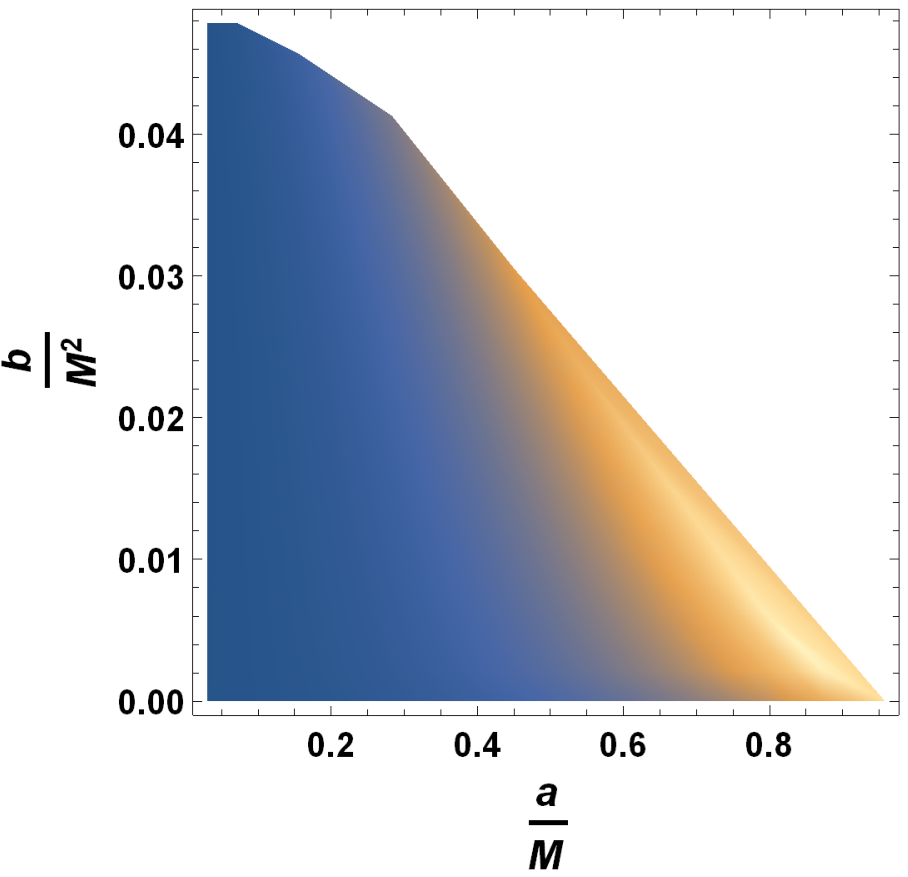}\hspace{1.5em}%
\end{subfigure}%
\begin{subfigure}{.28\textwidth}
\centering
\raisebox{.2\height}{\includegraphics[scale=.55]{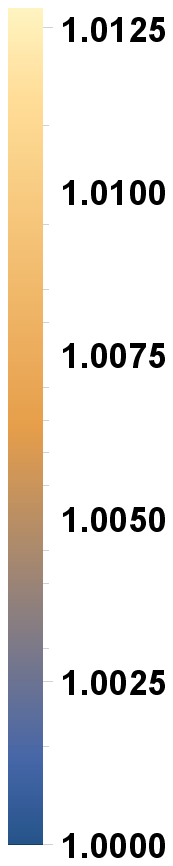}}\hspace{1.5em}%
\end{subfigure}%
\begin{subfigure}{.25\textwidth}
\centering
\includegraphics[scale=.65]{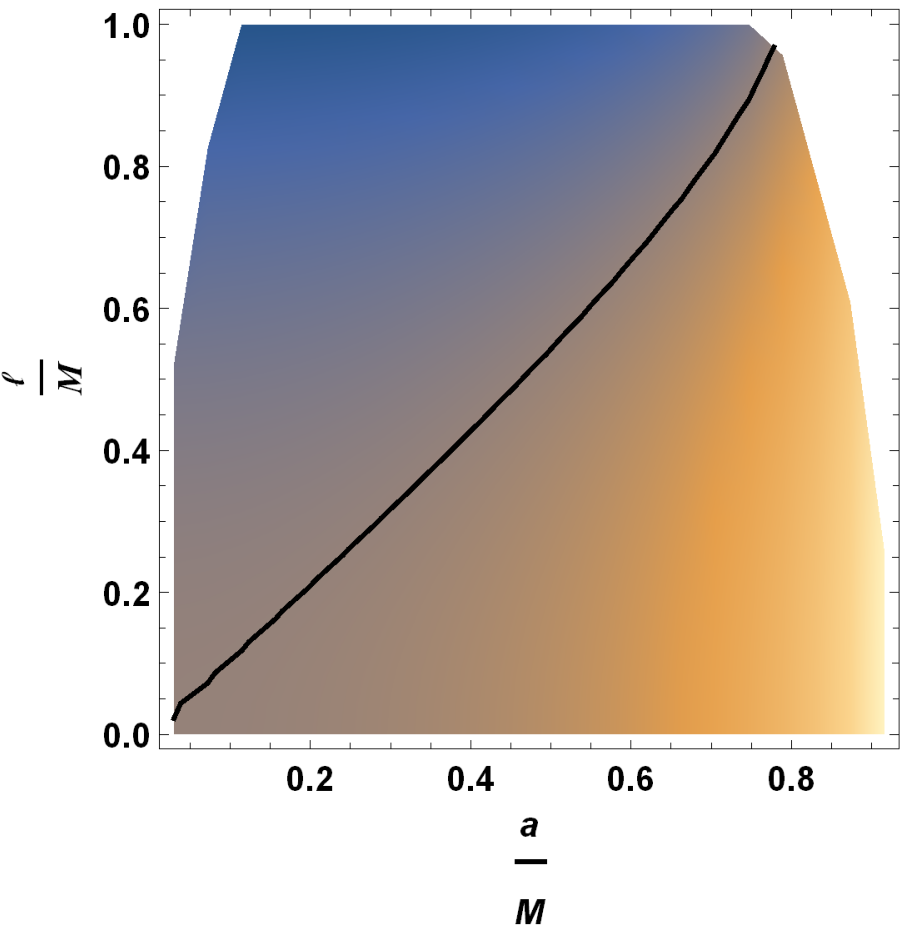}\hspace{1.5em}%
\end{subfigure}%
\begin{subfigure}{.28\textwidth}
\centering
\raisebox{.19\height}{\includegraphics[scale=.6]{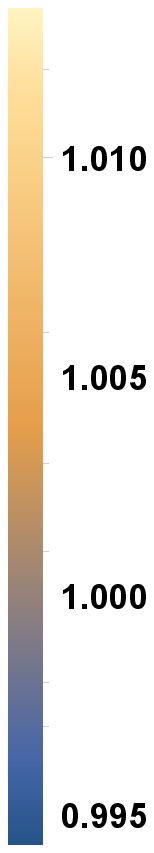}}
\end{subfigure}
\caption{The left panel is for $\ell=0.1M$ and the right panel is for
$b=0.001M^{2}$ where the inclination angle is $17^{o}$.}
\end{figure}
From the Figs. (22, 23), we observe that the condition $1 <D_{X}
\lesssim 4/3$ is satisfied for the finite parameter space of this
non-commutative black hole. Thus, non-commutative Simpson-Visser
black hole is remarkably consistent with EHT images of $M87^{*}$.
Therefore, observational data for the $M87^{*}$ black hole shadow can
not rule out the model designed for non-commutative Simpson-Visser
black holes.

It would be instructive to calculate the bound of the parameter
$\ell$ in a similar way we determined the bound of the Lorentz
violating parameter in the paper \cite{OUR} and of the non-commutative parameter $b$ in \cite{OUR1}. By modeling
M$87^{*}$ black hole as Kerr black hole, the author of the paper
\cite{RODRIGO} obtained a lower limit of $a$ for the M$87^{*}$
black hole. Bringing this result under consideration in \cite{OUR1}
we kept the interval of interest for $a$ as $[0.50M,0.99M]$, and
used the experimental constraints $\Delta C \leq 0.10$ and
$\theta_{d}=42\pm3 \mu as$. with these informations we observed that $b\in[0,0.000505973M^{2}]$
\cite{OUR1}. In a similar way, taking into account the bounds
$a\in[0.50M,0.99M]$ and $b\in[0,0.000505973M^{2}]$ and the
experimental constraints $\Delta C\leq0.1$ and
$\theta_{d}=39\pm3\mu as$, we get a bound on the parameter $\ell$. We
find that the parameter $\ell\in[0.827158M,0.85739801M]$. It is
intriguing to have upper and lower bounds of $\ell$ which are found
to be $0.827158M$ and $0.85739801M$ respectively. To the best of
our knowledge, the bound of the parameter $\ell$ from the shadow of the
astronomical black hole has not yet been reported so far.

\section{Summary and Conclusion}
In this paper, we have considered the black hole associated with
the Simpson-Visser spacetime background augmented with the
quantum correction due to the non-commutative aspect of spacetime.
Although Lorentz violation is inherent in the non-commutative
framework, the way this non-commutative correction is employed here the Lorentz symmetry remains undisturbed. Considering the quantum
corrections gained due to the non-commutativity of spacetime, we
study two important optical phenomena near this extended
Simpson-Visser background: superradiant scattering and shadow
cast. We begin our study with a description of the geometrical
aspects regarding the horizon and ergosphere of this modified
metric. We have thoroughly studied the cases when a black hole
exists and when it will represent a naked singularity for this
type of improved background. Graphically these are presented in
Fig.~(2) and we have sketched the ecospheres for different values of
the parameters associated with this modified spacetime background
in Fig.~(3).

After giving the outline of the geometry of the modified spacetime,
we move on to study the superradiance phenomena corresponding to
the black hole associated with this modified spacetime metric. We
observe that parameters $\ell$ and $b$ involved in the metric have
a pivotal role in the superradiance scattering along with its
dependence on the parameter $a$ linked with the spin of the black
hole. We observe that with the increase in the value of $a$ the
superradiance process intensifies. Also, the superradiance process
enhances with the increase in the value of $\ell$ and the reverse
is the case when the value of the parameters $a$ and $\ell$
decreases. We strikingly observe that with the increase in the
value of the parameter $b$ the superradiance process gets
diminished. The parameter $b$ is associated with the
non-commutative aspect of spacetime and it is considered a quantum
correction like the Lorentz violation parameter recently studied
in \cite{OUR1}. In the non-commutative Simpson-Visser black hole,
we notice the diminishing effect of quantum correction
superradiance phenomena. Note that the effect of quantum
correction in the superradiance associated with the violation of
Lorentz symmetry that set foot in the picture through the
bumblebee field was also of diminishing nature \cite{OUR1}

Through numerical simulation, we develop the shadow of this
modified black hole from the theoretical imputes associated with
the shadow. With the apprehension that the shadow of the $M87^*$
is equivalent to the shadow of this quantum corrected
Simpson-Visser black hole, we make an attempt to put constraint on
the value of the parameter $\ell$ involved in this
modified spacetime. We notice that the experimental constraint
$\Delta C \leq0.1$ is satisfied for finite parameter space when
the inclination angle is $90^{o}$, but for the inclination angle
$17^{o}$, the said experimental constrain is satisfied for the
entire parameter space. However, the constraint
$\theta_{d}=42\pm3\mu$ is found to satisfy a finite parameter
space within $1\sigma$ region for both the inclination angles
$\theta=90^{o}$ and $\theta=17^{o}$. Furthermore, we observe that
the measured axial ratio $4:3$ for $M87^{*}$ black hole is in good
agreement with $\Delta C \leq0.1$ for this setup.

It is intriguing if the bounds of the parameter
$\ell$ can be determined from the experimental observation of
astronomical black hole shadow. We have been able to estimate both
the upper and lower bounds of $\ell$ from our analysis. The upper
bound of $\ell$ is found to be $0.827158M$ and $0.85739801M$ is
found as the lower of $\ell$. To the best of our knowledge, the
estimation of the bond of the parameter $\ell$ from the astronomical
findings has not yet been reported so far. In \cite{OUR1}, we gave an estimate of the lower and upper
bounds of $b$. Here also, both the upper and lower bounds of $\ell$ have been possible to determine from the
findings of the astronomical black hole $M87^*$. It would be
attention-grabbing if it has been possible to estimate these
bounds from the observation of $SgrA^*$ and to what extent it
agrees with the present one.

\end{document}